\RequirePackage{ifpdf}
\ifpdf 
\documentclass[pdftex]{sigma}
\else
\documentclass{sigma}
\fi

\numberwithin{equation}{section}

\usepackage[mathscr]{eucal}

\newfont{\bbd}{msbm10 scaled\magstep1}

\def\ri{{\rm{i}}}

\begin{document}

\allowdisplaybreaks

\renewcommand{\thefootnote}{$\star$}

\renewcommand{\PaperNumber}{079}

\FirstPageHeading

\ShortArticleName{Linearizability of Nonlinear Equations on a Quad-Graph}

\ArticleName{Linearizability of Nonlinear Equations \\ on a Quad-Graph by a  Point, Two Points\\
 and Generalized Hopf--Cole Transformations\footnote{This
paper is a contribution to the Proceedings of the Conference ``Symmetries and Integrability of Dif\/ference Equations (SIDE-9)'' (June 14--18, 2010, Varna, Bulgaria). The full collection is available at \href{http://www.emis.de/journals/SIGMA/SIDE-9.html}{http://www.emis.de/journals/SIGMA/SIDE-9.html}}}

\Author{Decio LEVI~$^\dag$ and Christian SCIMITERNA~$^\ddag$}

\AuthorNameForHeading{D.~Levi and C.~Scimiterna}

\Address{$^\dag$~Dipartimento di Ingegneria Elettronica,   Universit\`a degli Studi Roma Tre and Sezione INFN,\\
\hphantom{$^\dag$}~Roma Tre,   Via della Vasca Navale 84, 00146 Roma, Italy}
\EmailD{\href{mailto:levi@roma3.infn.it}{levi@roma3.infn.it}}
\URLaddressD{\url{http://optow.ele.uniroma3.it/levi.html}}

\Address{$^\ddag$~Dipartimento di Fisica, Dipartimento di Ingegneria Elettronica,   Universit\`a degli Studi Roma\\
\hphantom{$^\ddag$}~Tre and Sezione INFN, Roma Tre,   Via della Vasca Navale 84, 00146 Roma, Italy}
\EmailD{\href{mailto:scimiterna@fis.uniroma3.it}{scimiterna@fis.uniroma3.it}}

\ArticleDates{Received April 15, 2011, in f\/inal form August 11, 2011;  Published online August 18, 2011}

\Abstract{In this paper we propose some  linearizability tests of partial dif\/ference equations on a quad-graph given by one point, two points and generalized Hopf--Cole transformations. We apply the so obtained tests to a set of nontrivial examples.}


\Classification{39A14}

\section{Introduction}

\begin{figure}[htbp]
\centering
\includegraphics{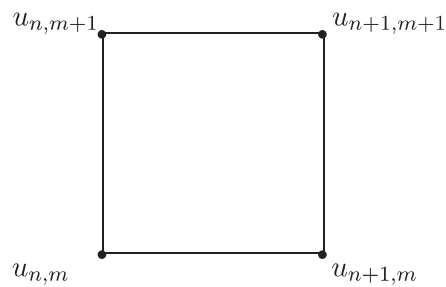}
\caption{The  quad-graph where a partial dif\/ference equation is def\/ined.}\label{fig1}
\end{figure}

In \cite{hls} one has provided necessary conditions for the linearizability of a vast class of real dispersive multilinear dif\/ference equations on a quad-graph (see Fig.~\ref{fig1}).
These conditions, obtained by considering the multiscale expansion up to f\/ifth order in the perturbation parameter, are not suf\/f\/icient to f\/ix all the free parameters of the equation and one needs to make some further ansatz or to use other techniques to f\/ix it.

So to verify these results and provide alternative ways to prove linearizability of partial dif\/ference equations on a quad-graph we write down here a set of algorithmic conditions obtained by postulating the existence of linearizing transformations, i.e.\ classes of transformations which  reduce a  given partial dif\/ference equation on a quad-graph
\begin{gather} \label{1.1a}
\mathcal E =\mathcal E (u_{n,m}, u_{n,m+1}, u_{n+1,m},u_{n+1,m+1} )=0, \qquad \frac{\partial \mathcal E}{\partial u_{n+i,m+j}}\not=0, \qquad i,j=0,1,
\end{gather}
for a f\/ield $u_{n,m}$ into a linear autonomous equation for $\tilde u_{n,m}$
\begin{gather} \label{1.3a}
{\tilde u}_{n,m}+a{\tilde u}_{n+1,m}+b{\tilde u}_{n,m+1}+c{\tilde u}_{n+1,m+1} = 0
\end{gather}
where $a$, $b$ and $c$ are some ($n,m$)-independent arbitrary nonzero constants. The choice that~(\ref{1.3a}) be autonomous  is a severe restriction but is a natural simplifying ansatz when one is dealing with autonomous equations. Moreover, as (\ref{1.1a}), (\ref{1.3a}) are taken to be autonomous  equations, i.e.\ they have no $n$, $m$ dependent coef\/f\/icients, they are translationally invariant under shifts in~$n$ and~$m$. So we can with no loss of generality choose as reference point $n=0$ and $m=0$. This will also be assumed to be  true for all the tranformations we will study in the following. We will consider here  one, two points and Hopf--Cole  transformations.

By a {\it one point transformation} we mean a transformation
\begin{gather} \label{1.3b}
\tilde u_{0,0} = f(u_{0,0})
\end{gather}
between (\ref{1.3a}) and (\ref{1.1a})  characterized by a function depending just from the function $u_{0,0}$ and maybe on some constant parameters. It will be a {\it Lie point transformation} if   $f=f_{0,0}$ satisf\/ies all Lie group axioms.   In the following we will only assume the dif\/ferentiability of the function~$f$ up to at least second order.

A natural generalization is when one considers {\it two points transformations}
\begin{gather} \label{1.3c}
\tilde u_{0,0} = g(u_{0,0},u_{0,1}),
\end{gather}
  characterized by a function $g=g_{0,0}$ depending on two lattice points,  $u_{0,0}$ and  $u_{0,1}$. The alternative choice when $g=g(u_{0,0},u_{1,0})$ will be studied elsewhere. Two lattice points is the minimum number of points necessary to provide in the continuous limit the f\/irst derivative and contact symmetries  have been introduced by Lie as  symmetries depending on f\/irst derivatives. Often contact symmetries are also called Miura transformations \cite{miura} as R.~Miura introduced them to transform the KdV into the  MKdV equation and have played a very important role in the integrability of the KdV equation.  Equation~(\ref{1.3c})  contains the transformation~(\ref{1.3b}) as a subcase but here we will assume $ \frac{\partial g}{\partial u_{0,1}} \ne 0$. Under this hypothesis  the conditions for point transformations are not obtained as a limiting case of the ones for contact transformations. So one and two points transformations will be treated as independent cases.

By a {\it generalized Hopf--Cole transformation} we mean a transformation \cite{cole,hopf}
\begin{gather} \label{1.3d}
 \tilde u_{0,1} = h(u_{0,0},u_{0,1}) \tilde u_{0,0}, \\ \label{1.3e}
 \tilde u_{1,0} = k(u_{0,0},u_{1,0},u_{0,1},u_{1,1}), \tilde u_{0,0}
\end{gather}
where $k$ is a scalar function given in term of $h$ by
\begin{gather} \label{1.3f}
k\doteq-\frac{1+bh_{0,0}(u_{0,0},u_{0,1})}{a+ch_{1,0}(u_{1,0},u_{1,1})}.
\end{gather}
The compatibility between (\ref{1.3d}) and (\ref{1.3e}) is to be identically satisf\/ied on the nonlinear equation~(\ref{1.1a}) while the condition~(\ref{1.3f}) is necessary for $\tilde u_{n,m}$ to satisfy the li\-near equation~(\ref{1.3a}). When $h(u_{0,0},u_{0,1})=u_{0,0}$  and $k(u_{0,0},u_{1,0},u_{0,1},u_{1,1})=u_{1,0} u_{0,0} - 2 u_{0,0}$ then $\tilde u_{0,0}$ will satisfy a~li\-near discrete heat equation and  $u_{0,0}$  a discrete Burgers equation~\cite{blr}. In such a case (\ref{1.3d}) is the discrete equivalent of the Hopf--Cole transformation which linearize the Burgers equation.

In Section~\ref{section2} we  discuss  point transformations, present the integrability conditions which ensure  that the given equation is a $C$-integrable equation and the dif\/ferential equations which def\/ine the function $f$. In a similar way in Section~\ref{section3} we analyze two points transformations. In  Section~\ref{section4} we present the  conditions which ensure  that the given equation is a $C$-integrable equation and the dif\/ferential equations which def\/ine the function $h$ for  the Hopf--Cole transformation.
In Section~\ref{section5} we shall present a few examples while in the f\/inal section we  present  some conclusive remarks and open problems.

\section{Linearization by a point transformation}\label{section2}

In this section we discuss  point transformations as, being def\/ined by a function of just one variable, they are the simplest transformation we can propose. We will state in detail the procedure used, which will be applied later in all the other cases.  This procedure follows a~similar one introduced in the case of the analysis of formal symmetries  for integrable quad-graph equations~\cite{ly09,ly11}. In particular we describe how we get from one side the determining equations which give the  transformation and on the other side the conditions under which the equation~(\ref{1.1a}) might be linearizable. The latter are  necessary conditions which the given equation has to sa\-tisfy if a~point transformation which linearizes the equation exists. If the conditions are satisf\/ied then we can solve the partial dif\/ferential equation determining the transformation and get a~f\/irst approxi\-ma\-tion to the point transformation. However only if also the initial determining equation is satisf\/ied the system is linearizable.

Assuming the existence of a point transformation (\ref{1.3b}) which linearizes (\ref{1.1a}), equation~(\ref{1.3a}) reads
\begin{gather} \label{3.4a}
f_{0,0}+af_{1,0}+bf_{0,1}+cf_{1,1}=0.
\end{gather}
In (\ref{3.4a}) and in the following equations we assume $u_{0j}$ and $u_{i0}$ as independent variables and consequently the variable $u_{1,1}$ appearing in the last term  is not independent but it can be written in term of independent variables using the equation~(\ref{1.1a})~\cite{ly09,mwx11}.  To be able to do so, we assume that (\ref{1.1a})
is solvable with respect to $u_{1,1}$
\begin{gather} \label{3.5a}
u_{1,1}=F (u_{0,0}, u_{0,1}, u_{1,0} ),
\end{gather}
where, as (\ref{1.1a}) depends on all lattice points we must have
\begin{gather} \label{3.6bb}
F_{,u_{0,1}}\not=0, \qquad F_{,u_{1,0}}\not=0, \qquad F_{,u_{0,0}}\not=0.
\end{gather}
To solve the functional equation (\ref{3.4a}) we apply the Abel technique \cite{abel}, i.e.\ we rewrite it as a~dif\/ferential equation. The solution of its dif\/ferential consequences is a necessary condition for the functional equation to be satisf\/ied.

Let us dif\/ferentiate (\ref{3.4a}) with respect to $u_{0,1}$ and then apply the logarithmic function. We get:
\begin{gather} \label{3.7a}
 (T_{1}-1 )\log\frac{d f_{0,1}} {d u_{0,1}}=\log\tilde F,
\end{gather}
where  $\tilde F$ is given by
\[
\tilde F\left(u_{0,0}, u_{0,1}, u_{1,0}\right)\doteq-\frac{b} {c F_{,u_{0,1}}}.
\]
We can always introduce a dif\/ferential operator $\mathcal A$ such that
\begin{gather} \label{3.7b}
\mathcal A\phi\left(u_{1,1}\right)=0,
\end{gather}
 where $\phi$ is an arbitrary function of its argument. The most general operator of this form reads
\[ 
  \mathcal A=\frac{\partial }{\partial u_{0,1}}+S^{(1)} (u_{0,1}, u_{0,0}, u_{1,0} )\frac{\partial }{\partial u_{1,0}}+S^{(2)} (u_{0,1}, u_{0,0}, u_{1,0} )\frac{\partial }{\partial u_{0,0}},
\]
where $S^{(i)} (u_{0,1}, u_{0,0}, u_{1,0} )$, $i=1,2$ are arbitrary functions of their  variables.
Equation (\ref{3.7b}) is satisf\/ied for any function $\phi$ if
\[ 
 S^{(1)}= -\frac{F_{,u_{0,1}}+S^{(2)}F_{,u_{0,0}}}{F_{,u_{1,0}}}.
\]
There is no further condition to f\/ix $S^{(2)} (u_{0,1}, u_{0,0}, u_{1,0} )$.
Applying the operator $\mathcal A$ onto  (\ref{3.7a}) we get
\begin{gather} \nonumber
-\frac{d} {d u_{0,1}}\log\frac{d f_{0,1}} {d u_{0,1}} = {\mathcal A}\log\tilde F \\
\label{3.11a}
 \hphantom{-\frac{d} {d u_{0,1}}\log\frac{d f_{0,1}} {d u_{0,1}}}{} =-\frac{1} {F_{,u_{1,0}}F_{,u_{0,1}}}\big(W_{(u_{0,1})} [F_{,u_{1,0}};F_{,u_{0,1}} ]+S^{(2)}W_{(u_{0,1})} [F_{,u_{1,0}};F_{,u_{0,0}} ]\big),
\end{gather}
which is a dif\/ferential equation for $f_{0,1}(u_{0,1})$.  In (\ref{3.11a}) $W_{(x)} [f;g ]$ is  the Wronskian operator with respect to the variable $x$ of the functions $f(x)$ and $g(x)$, def\/ined as
\[
W_{(x)} [f;g ]\doteq fg_{,x}-gf_{,x}.
\]
The left hand side of (\ref{3.11a}) depends only on $u_{0,1}$ while the right hand side depends on $u_{0,0}$, $u_{1,0}$ and $u_{0,1}$ through the given nonlinear dif\/ference equation and the up to now arbitrary function $S^{(2)} (u_{0,1}, u_{0,0}, u_{1,0} )$.  So  we must have
\[
\frac{\partial} {\partial u_{0,0}}{\mathcal A}\log\tilde F=0, \qquad \frac{\partial} {\partial u_{1,0}}{\mathcal A}\log\tilde F=0,
\]
i.e.
\begin{gather}
W_{(u_{0,1})} [F_{,u_{1,0}};F_{,u_{0,0}} ]S^{(2)}_{,u_{0,0}}+{\mathcal K}^{(1)}S^{(2)}+{\mathcal K}^{(0)}=0,\label{3.12a}\\
W_{(u_{0,1})} [F_{,u_{1,0}};F_{,u_{0,0}} ]S^{(2)}_{,u_{1,0}}+{\mathcal H}^{(1)}S^{(2)}+{\mathcal H}^{(0)}=0,\label{3.13a}
\end{gather}
with
\begin{gather*}
{\mathcal K}^{(0)} (u_{0,0},u_{1,0},u_{0,1} )\doteq\frac{W_{(u_{0,0})} [F_{,u_{1,0}}F_{,u_{0,1}};W_{(u_{0,1})} [F_{,u_{1,0}};F_{,u_{0,1}} ] ]}{F_{,u_{1,0}}F_{,u_{0,1}}},\\  \nonumber
{\mathcal K}^{(1)} (u_{0,0},u_{1,0},u_{0,1} )\doteq\frac{W_{(u_{0,0})} [F_{,u_{1,0}}F_{,u_{0,1}};W_{(u_{0,1})} [F_{,u_{1,0}};F_{,u_{0,0}} ] ]} {F_{,u_{1,0}}F_{,u_{0,1}}},\\  \nonumber
{\mathcal H}^{(0)} (u_{0,0},u_{1,0},u_{0,1} )\doteq\frac{W_{(u_{1,0})} [F_{,u_{1,0}}F_{,u_{0,1}};W_{(u_{0,1})} [F_{,u_{1,0}};F_{,u_{0,1}} ] ]}{F_{,u_{1,0}}F_{,u_{0,1}}},\\  \nonumber
{\mathcal H}^{(1)} (u_{0,0},u_{1,0},u_{0,1} )\doteq\frac{W_{(u_{1,0})} [F_{,u_{1,0}}F_{,u_{0,1}};W_{(u_{0,1})} [F_{,u_{1,0}};F_{,u_{0,0}} ] ]} {F_{,u_{1,0}}F_{,u_{0,1}}}.
\end{gather*}
As (\ref{3.12a}), (\ref{3.13a}) must be valid for any given function $S^{(2)} (u_{0,1}, u_{0,0}, u_{1,0} )$, it follows that
\begin{gather} \label{3.13b}
W_{(u_{0,1})} [F_{,u_{1,0}};F_{,u_{0,0}} ]=0
\end{gather}
and
\begin{gather} \label{3.13c}
{\mathcal K}^{(0)}=0, \qquad {\mathcal H}^{(0)}=0.
\end{gather}
When (\ref{3.13b}), (\ref{3.13c}) are satisf\/ied also ${\mathcal K}^{(1)}$ and $ {\mathcal H}^{(1)}$ are null. Equations~(\ref{3.13b}), (\ref{3.13c}) are necessary conditions for the linearizability by point transformations. If~(\ref{1.1a}) is multilinear, the condition ${\mathcal H}^{(0)}=0$ is always identically satisf\/ied.

In conclusion, we can state the following theorems:

\begin{theorem} \label{t1}
Given a partial difference equation \eqref{1.1a} on a quad-graph, there exists a linearizing point transformation only if \eqref{3.13b} and \eqref{3.13c} are  satisfied.
\end{theorem}

\begin{theorem} \label{t1a}
Given a partial difference equation \eqref{1.1a} on a quad-graph, if Theorem~{\rm \ref{t1}} is satisfied, then \eqref{3.11a} will provide the possible point transformation up to an integration constant. If the obtained $f(u_{0,0})$ verifies \eqref{3.4a} then the transformation \eqref{1.3b} linearizes our partial difference equation.
\end{theorem}

\section{Two points transformations}\label{section3}

Introducing the two points transformation (\ref{1.3c}) into the linear equation (\ref{1.3a}) we get
\begin{gather} \label{m1.3a}
g_{0,0}(u_{0,0},u_{0,1})+a g_{1,0}(u_{1,0},u_{1,1})+b g_{0,1}(u_{0,1},u_{0,2})+c g_{1,1}(u_{1,1},u_{1,2})=0
\end{gather}
an equation which has to be identically satisf\/ied for any value of the independent variables of our problem. In (\ref{m1.3a}), apart from the independent functions $u_{0,0}$, $u_{1,0}$, $u_{0,2}$ and $u_{0,1}$, there  appear the functions $u_{1,1}$ and $u_{1,2}$.  The function $u_{1,1}$ is expressed in term of the independent variables  taking into account the
 partial dif\/ference   quad-graph equation (\ref{1.1a}). An equation involving $u_{1,2}$ is obtained from (\ref{1.1a}) by shifting the second index by~1, i.e.\ considering  the
 partial dif\/ference  quad-graph equation $\mathcal E\left(u_{0,1}, u_{0,2}, u_{1,1},u_{1,2}\right)=0$. However this equation does not express directly~$u_{1,2}$ in terms of the independent variables but it involves $u_{1,1}$ which is expressed in term of the independent variables through the partial dif\/ference  quad-graph equation itself. This fact   complicates the  equations obtained for the dif\/ferential consequences of (\ref{m1.3a}). To avoid it we have two possibilities, either
back-shifting (\ref{1.3a}) once with respect to the second index, i.e.\ considering, in place of (\ref{m1.3a}), the following equation
\begin{gather} \label{3.4aa}
g_{0,-1}(u_{0,-1},u_{0,0})+a g_{1,-1}(u_{1,-1},u_{1,0})+b g_{0,0}(u_{0,0},u_{0,1})+c g_{1,0}(u_{1,0},u_{1,1}) = 0
\end{gather}
or, back-shifting once with respect to the f\/irst index and once with respect to the second, i.e.
\begin{gather}\label{3.4ab}
g_{-1,-1}(u_{-1,-1},u_{-1,0})\!+a g_{0,-1}(u_{0,-1},u_{0,0})\!+b g_{-1,0}(u_{-1,0},u_{-1,1})\!+c g_{0,0}(u_{0,0},u_{0,1}) = 0.\!\!\!
\end{gather}
In  (\ref{3.4aa}), (\ref{3.4ab})  the variables $u_{-1,-1}$, $u_{-1,1}$, $u_{1,-1}$ and $u_{1,1}$ are not independent and from (\ref{1.1a}) we have, apart from (\ref{3.5a})
\begin{gather*} 
u_{-1,1}=G(u_{-1,0}, u_{0,0}, u_{0,1}), \qquad
u_{-1,-1}=H(u_{0,0}, u_{-1,0}, u_{0,-1}), \\ u_{1,-1}=K(u_{0,-1}, u_{0,0}, u_{1,0}).
\end{gather*}
As   (\ref{1.1a}) depends on all lattice points we must have, apart from (\ref{3.6bb})
\begin{alignat*}{4} 
& G_{,u_{0,1}}\not=0, \qquad && G_{,u_{-1,0}}\not=0, \qquad && G_{,u_{0,0}}\not=0, & \nonumber \\
\nonumber
&H_{,u_{0,-1}}\not=0, \qquad && H_{,u_{-1,0}}\not=0, \qquad && H_{,u_{0,0}}\not=0,&  \\ \nonumber
& K_{,u_{0,-1}}\not=0, \qquad && K_{, u_{1,0}}\not=0, \qquad && K_{,u_{0,0}}\not=0. &
\end{alignat*}
Moreover, introducing the operators $T_1$ and $T_2$ such that $T_1 f_{0,0}=f_{1,0}$, $T_2 f_{0,0} = f_{0,1}$, we have the following relations between the derivatives of the functions $F$, $G$, $H$ and $K$:
\begin{alignat}{3}
&F_{,u_{0,1}} [T_1 G]_{,u_{1,1}} \bigl |_{u_{1,1}=F}=1, \qquad &&F_{,u_{1,0}} [T_2 K]_{,u_{1,1}}  \bigl |_{u_{1,1}=F} =1,& \nonumber \\
&K_{,u_{0,-1}} [T_1 H]_{,u_{1,-1}}  \bigl |_{u_{1,-1}=K} =1, \qquad &&G_{,u_{-1,0}} [T_2 H]_{,u_{-1,1}}  \bigl |_{u_{-1,1}=G}=1. & \label{3.6aab}
\end{alignat}
Taking into account (\ref{3.6aab}) it turns out that (\ref{3.4aa}) and (\ref{3.4ab}) give the same necessary conditions. So it is suf\/f\/icient to consider one of them, say (\ref{3.4aa}).

Dif\/ferentiating (\ref{3.4aa}) once with respect to $u_{0,1}$, we get
\begin{gather}\label{3.6aa}
b\frac{\partial g_{0,0}}{\partial u_{0,1}} ( u_{0,0},u_{0,1})+c\frac{\partial g_{1,0}}{\partial u_{1,1}}(u_{1,0}, u_{1,1})\frac{\partial F} {\partial u_{0,1}}( u_{0,0}, u_{0,1},u_{1,0})=0.
\end{gather}
Applying the logarithmic function to (\ref{3.6aa}) we get the dif\/ferential dif\/ference equation for the function $g$
\begin{gather}\label{3.6aaa}
 (T_{1}-1 )\log\frac{\partial g_{0,0}}{\partial u_{0,1}}=\log\tilde F, \qquad \tilde F( u_{0,0}, u_{0,1},u_{1,0})\doteq-\frac{1}{\frac{c}{b}\frac{\partial F}{\partial u_{0,1}}},
\end{gather}
where $\tilde F$ is an explicit function given in term of the given quad-graph partial dif\/ference equation~$\mathcal E$.
A solution of~(\ref{3.6aaa}) could be obtained by summing it up. However in this case the resulting solution $g$ would not be of the required form~(\ref{1.3c}).

So, to f\/ind a solution of (\ref{3.6aaa}) of the required form, we  simplify (\ref{3.6aaa}) by  introducing a dif\/fe\-ren\-tial operator $\mathcal A$ such that
\begin{gather} \label{3.7ab}
\mathcal A \phi (u_{1,0}, u_{1,1} )=0,
\end{gather}
where $\phi$ is an arbitrary function of its arguments. The most general operator of this form reads
\begin{gather}\label{3.7a'} 
\mathcal A=\frac{\partial }{\partial u_{0,1}}+S^{(1)}_{a} ( u_{0,0}, u_{0,1},u_{1,0} )\frac{\partial }{\partial u_{0,0}}+S^{(2)}_{a} ( u_{0,0}, u_{0,1},u_{1,0} )\frac{\partial }{\partial u_{1,0}},
\end{gather}
where $S^{(i)}_{a}\left( u_{0,0}, u_{0,1},u_{1,0}\right)$, $i=1,2$ are arbitrary functions of the independent variables to be determined. Equation~(\ref{3.7ab}) is satisf\/ied for any function $\phi$ if
\begin{gather}\label{3.7b'}
S^{(1)}_{a}=-\frac{F_{,u_{0,1}}}{F_{,u_{0,0}}}, \qquad S^{(2)}_{a}=0.
\end{gather}
Applying the operator $\mathcal A$ onto (\ref{3.6aaa}) and def\/ining $\psi (u_{0,0},u_{0,1} )\doteq \log\frac{\partial g_{0,0}}{\partial u_{0,1}}$, we get
\begin{gather}\label{3.8ac}
\psi_{,u_{0,1}}-\frac{F_{,u_{0,1}}}{F_{,u_{0,0}}}\psi_{,u_{0,0}}=\mathcal R (u_{0,0},u_{0,1},u_{1,0} )\doteq\frac{1}{F_{,u_{0,0}}F_{,u_{0,1}}}W_{(u_{0,1})}[F_{,u_{0,0}};F_{,u_{0,1}}].
\end{gather}
Equation~(\ref{3.8ac}) is a dif\/ferential equation for the function $\psi ( u_{0,0},u_{0,1} )$, i.e.\ for the function cha\-racterizing the two points transformation whose coef\/f\/icients depend on the given quad-graph partial dif\/ference equation~$\mathcal E$.  In~(\ref{3.8ac}) the function $\psi$ depends just on  $ u_{0,0}$, $u_{0,1}$ while the terms depending on the given quad-graph partial dif\/ference equation $\mathcal E$ depend on~$ u_{0,0}$,~$u_{0,1}$,~$u_{1,0}$.
As the quad-graph  equation~$\mathcal E$ depends also  on the  variable~$u_{1,0}$, (\ref{3.8ac}) will be an equation determining the two points transformation only if  some further compatibility conditions are satisf\/ied.

Dif\/ferentiating~(\ref{3.8ac}) once with respect to~$u_{1,0}$, we get the following  alternatives:

1.  If $W_{(u_{1,0})}[F_{,u_{0,1}};F_{,u_{0,0}}]=0$ identically, we must have
\begin{gather} \label{cond1}
\frac{\partial}{\partial u_{1,0}}\mathcal R=0,
\end{gather}
which is a necessary condition for the linearizability of (\ref{1.1a}) through the two points transformation (\ref{1.3c}).

2. If $W_{(u_{1,0})}[F_{,u_{0,1}};F_{,u_{0,0}}]\not=0$, we get
\begin{gather}\label{3.8ad}
\psi_{,u_{0,0}}=\mathcal M(u_{0,0},u_{0,1},u_{1,0}), \qquad \mathcal M\doteq\frac{F_{,u_{0,0}}^{2}\mathcal R_{,u_{1,0}}}{W_{(u_{1,0})}[F_{,u_{0,1}};F_{,u_{0,0}}]}.
\end{gather}
Inserting (\ref{3.8ad}) in (\ref{3.8ac}), we get
\begin{gather}\label{3.8ae}
\psi_{,u_{0,1}}=\mathcal N(u_{0,0},u_{0,1},u_{1,0}),
\end{gather}
where \[
\mathcal N\doteq\frac{F_{,u_{0,1}}}{F_{,u_{0,0}}}\mathcal M+\mathcal R.
\]
As the left hand side of  (\ref{3.8ad}) is independent of $u_{1,0}$, we get the necessary condition
\begin{gather} \label{3.8ag}
\frac{\partial}{\partial u_{1,0} } \mathcal M = 0.
\end{gather}
It is straightforward to prove that, if (\ref{3.8ag}) is satisf\/ied, also $\frac{\partial}{\partial u_{1,0}} \mathcal N = 0$ will be true.
Moreover  the compatibility of  (\ref{3.8ad}), (\ref{3.8ae}) gives another necessary condition
\begin{gather}\label{3.8af}
\frac{\partial}{\partial u_{0,1}}\mathcal M=\frac{\partial}{\partial u_{0,0}}\mathcal N.
\end{gather}

We can dif\/ferentiate  (\ref{3.4aa})  with respect to $u_{0,-1}$. In this case we get
\begin{gather} \label{3.6aaab}
 (T_{1}-1 )\log\frac{\partial g_{0,-1}}{\partial u_{0,-1}}=\log\tilde K, \qquad \tilde K(u_{0,0}, u_{1,0}, u_{0,-1})\doteq-\frac{1}{a\frac{\partial K}{\partial u_{0,-1}}}.
\end{gather}
We can introduce the dif\/ferential operator $\mathcal B$
\begin{gather}\label{3.14a'} 
\mathcal B=\frac{\partial }{\partial u_{0,-1}}+S^{(1)}_{b} (u_{0,0}, u_{1,0}, u_{0,-1} )\frac{\partial }{\partial u_{0,0}}+S^{(2)}_{b} (u_{0,0}, u_{1,0}, u_{0,-1} )\frac{\partial }{\partial u_{1,0}},
\end{gather}
 such that
\begin{gather} \label{3.7aba}
\mathcal B \xi (u_{1,0}, u_{1,-1} )=0,
\end{gather}
where $\xi$ is an arbitrary function of its arguments.
As   (\ref{3.7aba}) has to be satisf\/ied for any $\xi$, we get
\[
S^{(1)}_{b}=-\frac{K_{,u_{0,-1}}}{K_{,u_{0,0}}}, \qquad S^{(2)}_{b}=0.
\]
Applying the operator $\mathcal B$ onto (\ref{3.6aaab}) and def\/ining $\phi (u_{0,-1}, u_{0,0} )\doteq\log\frac{\partial g_{0,-1}}{\partial u_{0,-1}}$  we get the following dif\/ferential equation for the function $\phi$
\begin{gather}\label{3.8acc}
\phi_{,u_{0,-1}}-\frac{K_{,u_{0,-1}}}{K_{,u_{0,0}}}\phi_{,u_{0,0}}=\mathcal T (u_{0,0},u_{1,0},u_{0,-1} )\doteq\frac{1}{K_{,u_{0,0}}K_{,u_{0,-1}}}W_{(u_{0,-1})} [K_{,u_{0,0}};K_{,u_{0,-1}}].
\end{gather}

Dif\/ferentiating equation (\ref{3.8acc}) once with respect to $u_{1,0}$, we get the following  alternatives:

1.  $W_{(u_{1,0})}[K_{,u_{0,-1}};K_{,u_{0,0}}]=0$ identically, then we must have{\samepage
\begin{gather} \label{cond2}
\frac{\partial}{\partial u_{1,0}}\mathcal T=0,
\end{gather}
which is a necessary condition for linearizability of (\ref{1.1a}) through the two points transformation~(\ref{1.3c}).}

2. If $W_{(u_{1,0})}[K_{,u_{0,-1}};K_{,u_{0,0}}]\not=0$, we get
\begin{gather}\label{3.8adc}
\phi_{,u_{0,0}}=\mathcal P (u_{0,0},u_{1,0},u_{0,-1} ), \qquad \mathcal P\doteq\frac{K_{,u_{0,0}}^{2}\mathcal T_{,u_{1,0}}}{W_{(u_{1,0})} [K_{,u_{0,-1}};K_{,u_{0,0}}]}.
\end{gather}
Inserting (\ref{3.8adc}) in (\ref{3.8acc}), we get
\begin{gather}\label{3.8aec}
\phi_{,u_{0,-1}}=\mathcal Q (u_{0,0},u_{1,0},u_{0,-1} ),
\end{gather}
where
\[
\mathcal Q\doteq\frac{K_{,u_{0,-1}}}{K_{,u_{0,0}}}\mathcal P+\mathcal T.
\]
As the left hand side of  (\ref{3.8adc}) is independent of $u_{1,0}$, we get the necessary condition
\begin{gather}\label{3.8agcc}
\frac{\partial}{\partial u_{1,0}} \mathcal P =0.
\end{gather}
It is straightforward to prove that, if (\ref{3.8agcc}) is satisf\/ied, also $\frac{\partial}{\partial u_{1,0}} \mathcal Q=0$ will be true. Moreover  the compatibility of  (\ref{3.8adc}), (\ref{3.8aec}) gives another necessary condition
\begin{gather}\label{3.8afc}
\frac{\partial}{\partial u_{0,-1}}\mathcal P = \frac{\partial}{\partial u_{0,0}}\mathcal Q.
\end{gather}

We summarize the results so far obtained in the following theorems:

\begin{theorem}\label{t3}
Given a partial difference equation \eqref{1.1a} on a quad-graph we can construct the Wronskian functions $\mathbb F=W_{(u_{1,0})}[F_{,u_{0,1}};F_{,u_{0,0}}]$ and $\mathbb K=W_{(u_{1,0})}[K_{,u_{0,-1}};K_{,u_{0,0}}]$. Depending on the values of $\mathbb F$ and $\mathbb K$ we have different necessary conditions for the existence of a linearizing two points transformation \eqref{1.3c}.
\begin{enumerate}\itemsep=0pt
\item[$1.$]  $\mathbb F \not=0$,  $\mathbb K \not=0$.
We have  different results according to the value of $\mathcal M$.
\begin{enumerate}\itemsep=0pt
\item[$(a)$] If $\mathcal M\not=0$, apart from the linearizability  conditions \eqref{3.8ag},  \eqref{3.8af}, \eqref{3.8agcc}, \eqref{3.8afc} the following compatible conditions must be satisfied:
\begin{gather}\label{Maro}
\left(\log\frac{[T_{2}\mathcal P]}{\mathcal M}\right)_{,u_{0,0}}=\mathcal M-[T_{2}\mathcal Q], \qquad \left(\log\frac{[T_{2}\mathcal P]}{\mathcal M}\right)_{,u_{0,1}}=\mathcal N-[T_{2}\mathcal P].
\end{gather}
\item[$(b)$] If $\mathcal M=0$,  the linearizability conditions  are \eqref{3.8af}, \eqref{3.8agcc}, \eqref{3.8afc}.
\end{enumerate}

\item[$2.$]  $\mathbb F\not=0$,  $\mathbb K=0$.
Apart from the linearizability conditions   \eqref{3.8af}, \eqref{3.8ag}, \eqref{cond2} we have  different results according to the value of $\mathcal T_{,u_{0,0}}$.
\begin{enumerate}\itemsep=0pt
\item[$(a)$] If $\mathcal T_{,u_{0,0}}=0$
\[
\mathcal M_{,u_{0,0}}=\left(\frac{[T_{2}K]_{,u_{0,0}}}{[T_{2}K]_{,u_{0,1}}}\mathcal M\right)_{,u_{0,1}}+\left(\frac{[T_{2}K]_{,u_{0,0}}}{[T_{2}K]_{,u_{0,1}}}\mathcal N-\mathcal M+[T_2 \mathcal T]\right)\mathcal M.
\]
\item[$(b)$]  If $\mathcal T_{,u_{0,0}}\not=0$, defining
\[
\mathcal S\doteq-\frac{\left(\frac{[T_{2}K]_{,u_{0,0}}}{[T_{2}K]_{,u_{0,1}}}\mathcal M\right)_{,u_{0,1}}+\frac{[T_{2}K]_{,u_{0,0}}}{[T_{2}K]_{,u_{0,1}}}\mathcal M\mathcal N+[T_2 \mathcal T]\mathcal M-\mathcal M^2-\mathcal M_{,u_{0,0}}}{[T_2\mathcal  T]_{,u_{0,1}}},
\]
we get the following linearizability conditions
\[
\mathcal S_{,u_{0,1}}=\mathcal M-\mathcal S\mathcal N, \qquad \mathcal S_{,u_{0,0}}=\frac{[T_{2}K]_{,u_{0,0}}}{[T_{2}K]_{,u_{0,1}}}\mathcal M+[T_2 \mathcal T]\mathcal S-\mathcal S\mathcal M.
\]
\end{enumerate}

\item[$3.$]  $\mathbb F=0$,  $\mathbb K \not=0$.
Apart from the linearizability conditions    \eqref{3.8afc}, \eqref{3.8agcc}, \eqref{cond1} we have  different results according to the value of $\mathcal R_{u_{0,0}}$.
\begin{enumerate}\itemsep=0pt
\item[$(a)$]  If $\mathcal R_{,u_{0,0}}=0$,
\[
\left([T_{2}\mathcal P]\right)_{,u_{0,1}}=\left(\frac{F_{,u_{0,1}}}{F_{,u_{0,0}}}[T_{2}\mathcal P]\right)_{,u_{0,0}}+\left(\frac{F_{,u_{0,1}}}{F_{,u_{0,0}}}[T_{2}\mathcal Q]-[T_{2}\mathcal P]+\mathcal R\right)[T_{2}\mathcal P].
\]
\item[$(b)$]  If $\mathcal R_{,u_{0,0}}\not=0$, defining
\begin{gather*} 
\mathcal U \doteq -\frac{\left(\frac{F_{,u_{0,1}}}{F_{,u_{0,0}}}[T_{2}\mathcal P]\right)_{,u_{0,0}}+\frac{F_{,u_{0,1}}}{F_{,u_{0,0}}}[T_{2}\mathcal P ][T_{2}\mathcal Q]+\mathcal R [T_{2}\mathcal P]-([T_{2}\mathcal P])^2-([T_{2}\mathcal P])_{,u_{0,1}}}{\mathcal R_{,u_{0,0}}}
\end{gather*}
we get the further linearizability conditions
\[
\mathcal U_{,u_{0,0}}=[T_{2}\mathcal P]-\mathcal U [T_{2}\mathcal Q], \qquad \mathcal U_{,u_{0,1}}=\frac{F_{,u_{0,1}}}{F_{,u_{0,0}}}[T_{2}\mathcal P]+\mathcal R\mathcal U-\mathcal U [T_{2}\mathcal P].
\]
\end{enumerate}

\item[$4.$] $\mathbb F=0$,  $\mathbb K=0$.
Apart from the linearizability conditions    \eqref{cond2}, \eqref{cond1} we have  a set of conditions for the functions  $F$ and $K$ involved, depending  if $\frac{[T_2 K]_{u_{0,0}}}{[T_2 K]_{u_{0,1}}}\frac{ F_{u_{0,1}}}{F_{u_{0,0}}}$ is equal to $1$ or not. These conditions are obtained by requiring that  the overdetermined system obtained by explicitating  \eqref{3.8ac}, \eqref{3.8acc} in term of $g=g_{0,0}$ and possibly shifting
\begin{gather} \label{3.8amd}
 g_{,u_{0,0},u_{0,0}}-\frac{[T_2 K]_{,u_{0,0}}}{[T_2 K]_{,u_{0,1}}} g_{,u_{0,0},u_{0,1}}  =[T_2 \mathcal T] g_{,u_{0,0}}, \\ \label{3.8ame}
 g_{,u_{0,1},u_{0,1}} - \frac{F_{,u_{0,1}}}{F_{,u_{0,0}}} g_{,u_{0,0},u_{0,1}}  =\mathcal R g_{,u_{0,1}},
\end{gather}
be solvable for any $u_{0,0}$, $u_{0,1}$, $u_{1,0}$.
These equations are easy to derive by symbolic mani\-pulation but too long to write down. So, for the sake of clarity, we do not write them down here.
\end{enumerate}
\end{theorem}

\begin{theorem} \label{t3a}
Given a partial difference equation \eqref{1.1a} on a quad-graph, if Theorem {\rm \ref{t3}} is satisfied,   depending on the values of $\mathbb F$ and $\mathbb K$, we have different partial differential equations defining the two points transformation.
\begin{enumerate}\itemsep=0pt
\item[$1.$] $\mathbb F \ne 0$,  $\mathbb K \ne 0$.
We have  different results according to the value of $\mathcal M$.
\begin{enumerate}\itemsep=0pt
\item[$(a)$] If $\mathcal M\not=0$ we have  for $g=g_{0,0}$
\[
g_{,u_{0,0},u_{0,0}}=[T_{2}\mathcal Q] g_{,u_{0,0}}, \qquad g_{u_{0,1}}=\frac{[T_{2}\mathcal P]}{\mathcal M}g_{,u_{0,0}}.
\]
\item[$(b)$] If $\mathcal M=0$ we have   $g=g^{(0)}(u_{0,0}) + g^{(1)}(u_{0,1})$  and
\[
g^{(0)}_{,u_{0,0} u_{0,0}} = [T_2 \mathcal Q] g^{(0)}_{,u_{0,0}}, \qquad g^{(1)}_{,u_{0,1} u_{0,1}} =  \mathcal N g^{(1)}_{,u_{0,1}}.
\]
\end{enumerate}\itemsep=0pt
\item[$2.$] $\mathbb F \ne 0$,  $\mathbb K=0$.
We have  different results according to the value of $\mathcal T_{u_{0,0}}$.
\begin{enumerate}
\item[$(a)$] If $\mathcal T_{,u_{0,0}}=0$ the two points transformation is obtained by solving   for $g=g_{0,0}$ the compatible system of  partial differential equations \eqref{3.8ad}, \eqref{3.8ae} and \eqref{3.8amd};
\item[$(b)$] If $\mathcal T_{,u_{0,0}} \ne 0$ the two points transformation is obtained by solving the compatible  partial differential equations \eqref{3.8ae} and
\[ 
g_{,u_{0,0}}=\mathcal S   g_{,u_{0,1}}.
\]
\end{enumerate}
\item[$3.$] $\mathbb F=0$,  $\mathbb K \ne 0$.
We have  different results according to the value of $\mathcal R_{,u_{0,0}}$.
\begin{enumerate}\itemsep=0pt
\item[$(a)$] If $\mathcal R_{,u_{0,0}}=0$ the two points transformation is obtained by solving   for $g=g_{0,0}$ the compatible system of  partial differential equations \eqref{3.8adc}, \eqref{3.8aec} and \eqref{3.8ame};
\item[$(b)$] If $\mathcal R_{,u_{0,0}} \ne 0$ the two points transformation is obtained by solving the compatible  partial differential equations~\eqref{3.8aec} and
\[ 
g_{,u_{0,1}} = \mathcal U   g_{,u_{0,0}}.
\]
\end{enumerate}
\item[$4.$] $\mathbb F=0$,  $\mathbb K=0$ the two points transformation is obtained by solving   for $g=g_{0,0}$ the compatible system of  partial differential equations \eqref{3.8ame} and \eqref{3.8amd}.
\end{enumerate}
\end{theorem}

When we have solved the PDEs for the function $g(u_{0,0},u_{0,1})$ we may still have arbitrary functions or arbitrary constants. These get f\/ixed by inserting the function $g$ into the equations~(\ref{3.6aa}),~(\ref{3.6aaab}) and solving them and their consequences. At the end we need to verify that~(\ref{m1.3a}) or its shifted versions~(\ref{3.4aa}),~(\ref{3.4ab}) be satisf\/ied.

\section[Linearization by a generalized Hopf-Cole transformation]{Linearization by a generalized Hopf--Cole transformation} 
\label{section4}

Let us assume the existence of a generalized Hopf--Cole transformation (\ref{1.3d}) which linearizes~(\ref{1.1a}) into~(\ref{1.3a}),
where $a$, $b$ and $c$ are arbitrary constants and the function $h=h_{0,0}$ of its two arguments is to be determined.

The nonlinear equation (\ref{1.1a}) can be written just in term of $h$ and $k$ as
\begin{gather}\label{3.6}
h_{1,0} k_{0,0} = k_{0,1} h_{0,0}
\end{gather}
and in terms of $h$ alone as
\begin{gather}\label{3.7}
\left(\frac{h_{0,0}+1/b}{h_{0,0}}\right) = \left( \frac{h_{1,0}+a/c}{h_{1,0}} \right) \left( \frac{h_{0,1}+1/b}{h_{1,1}+a/c} \right).
\end{gather}

If we assume $h=h(u_{0,0})$, then, dif\/ferentiating ({\ref{3.7}}) with respect to $u_{0,1}$, we f\/ind
\begin{gather} \label{Amaryllis}
\frac{\partial }{\partial u_{0,1}} \log (h_{0,1} + 1/b)=\frac{\partial}{\partial u_{1,1}}\log (h_{1,1} + a/c)\frac{\partial}{\partial u_{0,1}} F,
\end{gather}
and, by carrying out the same kind of calculations as in Section~\ref{section2}, we f\/ind the same linearizability conditions (\ref{3.13b}), (\ref{3.13c}) as for point transformations, i.e.\ Theorem~\ref{t1} will be valid. However  the dif\/ferential equation for the transformation is dif\/ferent and is given by
\begin{gather} \label{Reghini}
\frac{d} {d u_{0,1}}\log\frac{d} {d u_{0,1}}\log (h_{0,1}+1/b )=\frac{1} {F_{,u_{1,0}}F_{,u_{0,1}}}W_{(u_{0,1})} [F_{,u_{1,0}};F_{,u_{0,1}} ].
\end{gather}
So, 
if an equation is linearizable by a point transformation it can also be linearizable by a~Hopf--Cole transformation depending on one point only. However the ef\/fective linearizing transformation is dif\/ferent and thus one can f\/ind a~non\-linear partial dif\/ference equation on the square which is linearizable by a Hopf--Cole transformation with $h=h(u_{0,0})$ but not by a point transformation~(\ref{1.3b}).

If the left hand side of (\ref{3.7}) depends on $u_{0,0}$ and $u_{0,1}$, the f\/irst left term in the right hand side depends on  $u_{1,0}$ and $u_{1,1}$ and the second one on $u_{0,1}$, $u_{0,2}$, $u_{1,1}$ and $u_{1,2}$. The variable $u_{1,1}$ is given in terms of the independent variables by (\ref{3.5a}) while $u_{1,2}$ can be rewritten  in term of the independent  variables as
\begin{gather} \label{3.8}
 u_{1,2}=F(u_{0,1}, u_{0,2}, F(u_{0,0}, u_{0,1}, u_{1,0})).
\end{gather}
So, as from (\ref{3.8})  the expression of $u_{1,2}$ in terms of the independent variables depends twice on the quad-graph equation (\ref{1.1a}),  we will consider in place of (\ref{3.6}) the equation,
\[ 
  h_{1,-1} k_{0,-1} = k_{0,0} h_{0,-1}.
\]
which in terms of $h$ alone read
\begin{gather} \label{3.10}
\left(\frac{h_{0,-1}+1/b}{h_{0,-1}}\right)=\left(\frac{h_{1,-1}+a/c}{h_{1,-1}}\right)\left(\frac{h_{0,0}+1/b}{h_{1,0}+a/c} \right),
\end{gather}
The left hand side of (\ref{3.10}) depends on $u_{0,0}$ and $u_{0,-1}$, the f\/irst left term in the right hand side depends on  $u_{1,0}$ and $u_{1,-1}=K(u_{0,-1}, u_{1,0}, u_{0,0})$ and the second one on $u_{0,1}$, $u_{0,0}$, $u_{1,1}=F(u_{0,1}, u_{1,0}, u_{0,0})$ and $u_{1,0}$.
 So, the term on the left hand side of (\ref{3.10}) depends on $u_{0,0}$ and $u_{0,-1}$, the f\/irst left term on the right hand side depends on  $u_{1,0}$, $u_{0,0}$ and $u_{0,-1}$ while the second one on $u_{0,1}$, $u_{0,0}$ and $u_{1,0}$.
Thus one can see that the three terms appearing in the equation (\ref{3.10}) contain no overlapping set of variables. This is a  condition necessary to get out of (\ref{3.10}) some dif\/ferential conditions for the functions $F$ and $K$,  i.e.\ for the equation (\ref{1.1a}) to be rewritable as the compatibility condition of (\ref{1.3d}) and (\ref{1.3e}).

Let us consider (\ref{3.10}) and, as we have products, we  reduce it to a sum of terms by applying to it the logarithmic function. Then we dif\/ferentiate the resulting equation with respect to $u_{0,1}$. Only the second term on the r.h.s.\ of the equality depends on $u_{0,1}$ through the dependence of~$h_{1,0}$ on~$u_{1,1}$ and of~$h_{0,0}$. So we get:
\begin{gather} \label{3.14}
\frac{\partial }{\partial u_{0,1}} \log (h_{0,0} + 1/b)=\frac{\partial }{\partial u_{1,1}}\log (h_{1,0} + a/c) \frac{\partial}{\partial u_{0,1}} F,
\end{gather}
equivalent, in structure to (\ref{3.6aa}).
The term on the l.h.s.\ of (\ref{3.14}) depends on $u_{0,0}$ and $u_{0,1}$ while the f\/irst factor on the r.h.s.\ depends on $u_{1,0}$ and $u_{1,1}$ and we can always
introduce the differential operator $\mathcal A$  as given by~(\ref{3.7a'}).
So, if we apply again the logarithmic function to equation~(\ref{3.14}) and then the operator $\mathcal A$ onto the resulting equation, setting $\psi (u_{0,0},u_{0,1} )\doteq\log\frac{\partial}{\partial u_{0,1}}\log (h+1/b )$, we get the linear dif\/ferential equation
 (\ref{3.8ac}). It is worthwhile to notice that, even if the dif\/ferential equation is the same when expressed in term of the variable $\psi$, its expression in term of~$f$ is dif\/ferent from the one in term of~$h$.

Let us now dif\/ferentiate (\ref{3.10}) with respect to $u_{0,-1}$. Proceeding in an analogous way as we did before, we get
\begin{gather} \label{3.14a}
\frac{\partial}{\partial u_{0,-1}} \log \left(\frac{h_{0,-1} + 1/b}{h_{0,-1}}\right)=\frac{\partial }{\partial u_{1,-1}} \log \left(\frac{h_{1,-1} + a/c}{h_{1,-1}}\right) \frac{\partial }{\partial u_{0,-1}} K.
\end{gather}
The term on the l.h.s.\ of (\ref{3.14a}) depends on $u_{0,-1}$ and $u_{0,0}$ while the f\/irst factor on the r.h.s.\ depends on $u_{1,-1}$ and $u_{1,0}$.
We can always
introduce the differential operator~$\mathcal B$  as given by~(\ref{3.14a'}).
So if we apply again the logarithmic function to equation (\ref{3.14a}) and then the operator $\mathcal B$ onto the resulting equation, setting $\phi (u_{0,-1},u_{0,0} )\doteq\log\frac{\partial}{\partial u_{0,-1}}\log\frac{h_{0,-1}+1/b}{h_{0,-1}}$, we get the linear dif\/ferential equation (\ref{3.8acc}) for $\phi$. However, as before, even if the dif\/ferential equation is the same when expressed in term of the variable~$\phi$, its expression in term of~$f$ is dif\/ferent from the one in term of $h$.

As the determining equations in terms of $\psi$ and $\phi$ are exactly the same as those  of contact transformations, the linearizability conditions are as presented in Theorem~\ref{t3}. However the function $\psi$ and $\phi$ are def\/ined here dif\/ferently then in the case of two points transformations. So the equations def\/ining $h=h_{0,0}$ are dif\/ferent. In particular (\ref{3.8amd}) and (\ref{3.8ame}) in this case have to be replaced by the nonlinear equations
\begin{gather*} 
 \frac{h_{u_{0,1} u_{0,1}}}{h+1/b}-\left(\frac{h_{u_{0,1}}}{h+1/b} \right)^2
- \frac{F_{u_{0,1}}}{F_{u_{0,0}}} \left(\frac{h_{u_{0,1} u_{0,0}}}{h+1/b}-\frac{h_{u_{0,1}} h_{u_{0,0}}}{(h+1/b)^2} \right) = \mathcal R, \\ \nonumber 
 \left[\left( \frac{h_{0,-1,u_{0,-1}}}{h_{0,-1}(h_{0,-1}+1/b)} \right)_{u_{0,-1}} -\frac{K_{u_{0,-1}}}{K_{u_{0,0}}} \left(\frac{h_{0,-1,u_{0,-1}}}{h_{0,-1}(h_{0,-1}+1/b)} \right)_{u_{0,0}}\right] \frac{h_{0,-1}(h_{0,-1}+1/b)}{h_{0,-1,u_{0,-1}}} = \mathcal T.
\end{gather*}
A simpler and sometimes more useful nonlinear equation for $h_{0,0}$ can be obtained in the following way. Let us shift (\ref{3.14a}) by $T_2$. In such a way we  get
\begin{gather} \label{3.14b}
\frac{\partial }{\partial u_{0,0}} \log \left(\frac{h_{0,0} + 1/b}{h_{0,0}}\right) = \frac{\partial }{\partial u_{1,0}} \log \left(\frac{h_{1,0} + a/c}{h_{1,0}}\right) \frac{\partial }{\partial u_{0,0}} [T_2 K].
\end{gather}
Then  from (\ref{3.14}), (\ref{3.14b}) we  extract the partial derivatives of $h_{0,0}$ with respect to $u_{0,0}$ and $u_{0,1}$,
\begin{gather} \label{3.14c}
h_{0,0,u_{0,1}}  =  \frac{h_{0,0}+1/b}{h_{1,0}+a/c} h_{1,0,u_{1,1}} \frac{\partial F}{\partial u_{0,1}}, \\ \label{3.14d}
h_{0,0,u_{0,0}}  = \frac{a/c h_{0,0}}{1/b h_{1,0}} \frac{h_{0,0}+1/b}{h_{1,0}+a/c} h_{1,0,u_{1,0}} \frac{\partial [T_2 K]}{\partial u_{0,0}}.
\end{gather}
Dividing (\ref{3.14c}) by (\ref{3.14d}) we get the following  equation:
\begin{gather} \label{3.25}
(T_1-1) \log \frac{h_{0,0} h_{0,0,u_{0,1}}}{h_{0,0,u_{0,0}}} + \log \frac{1/b \frac{\partial F}{\partial u_{0,1}}}{a/c  \frac{\partial [T_2 K]}{\partial u_{0,0}}} = 0.
\end{gather}
Dif\/ferentiating (\ref{3.25}) with respect to $u_{1,0}$ we obtain a second order nonlinear dif\/ferential equation for the function $\chi(u_{0,0}, u_{0,1})= \log(h_{0,0})$.
We have:
\begin{gather} \label{3.26}
\left[\chi_{,u_{0,0}} + \frac{\chi_{,u_{0,0},u_{0,1}} }{\chi_{,u_{0,1}} } - \frac{\chi_{,u_{0,0},u_{0,0}} }{\chi_{,u_{0,0}} } \right] +
\mathcal C\left[ \chi_{,u_{0,1}} + \frac{\chi_{,u_{0,1},u_{0,1}} }{\chi_{,u_{0,1}} } - \frac{\chi_{,u_{0,1},u_{0,0}} }{\chi_{,u_{0,0}} } \right ]+\mathcal D=0,
\end{gather}
where
\begin{gather*} 
\mathcal C (u_{-1,0},u_{0,0},u_{0,1} )\doteq\left\{T_{1}^{-1} \left[\frac{\partial F}{\partial u_{1,0}}\right]\right\}_{u_{-1,1}\rightarrow G (u_{-1,0},u_{0,0},u_{0,1} )},\\\nonumber
\mathcal D (u_{-1,0},u_{0,0},u_{0,1} )\doteq \left\{T_1^{-1} \left[\frac{\partial}{\partial u_{1,0}}\log\left(\frac{1}{\mathcal H}\frac{\partial F}{\partial u_{0,1}}\right)\right]\right\}_{u_{-1,1}\rightarrow G (u_{-1,0},u_{0,0},u_{0,1} )},\\\nonumber
\mathcal H (u_{0,0},u_{1,0},u_{0,1} )\doteq\left\{T_{2}\left[\frac{\partial K}{\partial u_{0,-1}}\right]\right\}_{u_{1,1}\rightarrow F (u_{0,0},u_{1,0},u_{0,1} )}.
\end{gather*}
The f\/irst and second terms of (\ref{3.26}) depend  on derivatives of the unknown function $\chi(u_{0,0},u_{0,1})$ but the coef\/f\/icient of the second term and the last one may contain also~$u_{-1,0}$. So we have a~further set of linearizability conditions.
If $\frac{\partial}{\partial u_{-1,0}} \mathcal C =0$, dif\/ferentiating (\ref{3.26}) with respect to~$u_{-1,0}$ we  have
\[ 
\frac{\partial}{\partial u_{-1,0}} \mathcal D =0,
\]
while, if $\frac{\partial}{\partial u_{-1,0}} \mathcal C \ne 0$, after a dif\/ferentiation with respect to $u_{-1,0}$, we  have
\begin{gather} \label{3.29}
W_{(u_{-1,0})} [\mathcal C_{,u_{-1,0}};\mathcal D_{,u_{-1,0}} ] =0.
\end{gather}
In the f\/irst case, the solutions of (\ref{3.26}) provides us with an ansatz of the function $h$, otherwise the function $h$ is obtained by solving the following overdetermined system of nonlinear partial dif\/ferential equations
\begin{gather} \label{3.30}
\chi_{,u_{0,0}}  =  \frac{\chi_{,u_{0,0},u_{0,0}} }{\chi_{,u_{0,0}} }-\frac{\chi_{,u_{0,0},u_{0,1}} }{\chi_{,u_{0,1}} }+\frac{W_{(u_{-1,0})}  [\mathcal C;\mathcal D  ]} {\mathcal C_{,u_{-1,0}}},\\\nonumber
\chi_{,u_{0,1}}  = \frac{\chi_{,u_{0,1},u_{0,0}} }{\chi_{,u_{0,0}} }-\frac{\chi_{,u_{0,1},u_{0,1}} }{\chi_{,u_{0,1}}}-\frac{\mathcal D_{,u_{-1,0}}}{\mathcal C_{,u_{-1,0}}}.
\end{gather}
If the condition (\ref{3.29}) is satisf\/ied, then $\frac{\partial}{\partial u_{-1,0}}\frac{W_{(u_{-1,0})}  [\mathcal C;\mathcal D  ]} {\mathcal C_{,u_{-1,0}}}=0$. The overdetermined system~(\ref{3.30}) is compatible if\/f
\begin{gather} \label{3.31}
W_{(u_{0,1})}   [W_{(u_{-1,0})}   [\mathcal C;\mathcal D  ];\mathcal C_{,u_{-1,0}}   ]=W_{(u_{0,0})}   [\mathcal C_{,u_{-1,0}};\mathcal D_{,u_{-1,0}}  ],
\end{gather}
a further linearizability condition. Equations (\ref{3.26}), (\ref{3.30}) are a nonlinear partial dif\/ferential system which, introducing the function
\begin{gather}\label{San3}
\theta\left(u_{0,0},u_{0,1}\right)\doteq\chi+\log\left(\frac{\chi_{,u_{0,1}}}{\chi_{,u_{0,0}}}\right),
\end{gather}
can be linearized and read:
\begin{gather}\label{San1}
\theta_{,u_{0,0}}+\mathcal C\theta_{,u_{0,1}}+\mathcal D=0,
\\
\label{San2}
\theta_{,u_{0,0}}=\frac{W_{(u_{-1,0})}   [\mathcal C;\mathcal D  ]} {\mathcal C_{,u_{-1,0}}},\qquad\theta_{,u_{0,1}}=-\frac{\mathcal D_{,u_{-1,0}}}{\mathcal C_{,u_{-1,0}}}.
\end{gather}
Once the solution of the equations (\ref{San1}) or (\ref{San2}) has been obtained, the function $h$ can be reconstructed. Starting from the def\/inition (\ref{San3}) we  get
\[
e^{\chi}\frac{\partial}{\partial u_{0,1}} e^{\chi}=e^{\theta}\frac{\partial}{\partial u_{0,0}} e^{\chi},
\]
or in terms of $h$
\begin{gather}\label{San4}
h_{,u_{0,0}}=e^{-\theta}h h_{,u_{0,1}},
\end{gather}
which is a \emph{Hopf-like} equation whose solution can be obtained for example by separation of variables. Once we have a solution, we can introduce it into the lowest order dif\/ferential equations and def\/ine
the arbitrary functions or constant involved. The so obtained function $h$ will provide us with a linearizing generalized Hopf--Cole transformation if the dif\/ference relation (\ref{3.7}) is satisf\/ied.
Equation~(\ref{San4}) can be introduced in (\ref{3.14c}), (\ref{3.14d}) and after some manipulations and the application of the operator $\mathcal A$ def\/ined in (\ref{3.7a'}), (\ref{3.7b'}), we obtain a linear evolution equation for the function $\theta (u_{0,0},u_{0,1} )$
\begin{gather}\label{San5}
\theta_{,u_{0,1}}-\frac{F_{,u_{0,1}}}{F_{,u_{0,0}}}\theta_{,u_{0,0}}=\tilde {\mathcal T}  (u_{0,0},u_{1,0},u_{0,1} )\doteq\mathcal A\log\left(\frac{1}{\mathcal H} F_{,u_{0,1}}\right).
\end{gather}

Dif\/ferentiating equation~(\ref{San5}) once with respect to $u_{1,0}$, we get the following alternatives:

1. $\mathbb F=0$ identically, then we must have
\[ 
\tilde {\mathcal T}_{u_{1,0}}=0,
\]
which is a necessary condition for linearizability through the Hopf--Cole transformation (\ref{1.3d}), (\ref{1.3e}), (\ref{1.3f}).

2. If $\mathbb F\not=0$, we get
\begin{gather}\label{44}
\theta_{,u_{0,0}}=\tilde {\mathcal P} (u_{0,0},u_{1,0},u_{0,1} ), \qquad \tilde {\mathcal P}\doteq\frac{F_{,u_{0,0}}^{2}\tilde {\mathcal T}_{,u_{1,0}}}{\mathbb F}.
\end{gather}
Inserting (\ref{44}) in (\ref{San5}), we get
\begin{gather}\label{45}
\theta_{,u_{0,1}}=\tilde {\mathcal Q} (u_{0,0},u_{1,0},u_{0,1} ),
\end{gather}
where
\[
\tilde {\mathcal Q}\doteq\frac{F_{,u_{0,1}}}{F_{,u_{0,0}}}\tilde {\mathcal P}+\tilde {\mathcal T}.
\]
As the left hand side of  (\ref{44}) is independent of $u_{1,0}$, we get the necessary condition
\begin{gather}\label{46}
\frac{\partial}{\partial u_{1,0}} \tilde {\mathcal P} =0.
\end{gather}
It is straightforward to prove that, if (\ref{46}) is satisf\/ied, also $\frac{\partial}{\partial u_{1,0}} \tilde {\mathcal Q}=0$ will be true. Moreover  the compatibility of  (\ref{44}), (\ref{45}) gives another necessary condition
\[ 
\frac{\partial}{\partial u_{0,1}}\tilde {\mathcal P} = \frac{\partial}{\partial u_{0,0}}\tilde {\mathcal Q}.
\]

As in the case of contact transformations, the combination of the two cases def\/ined by (\ref{San1}) or (\ref{San2}) and the two cases def\/ined by  (\ref{San5}) or (\ref{44}), (\ref{45}) gives a total of four subcases for the specif\/ication of the function $\theta$. If the conditions (\ref{3.29}), (\ref{3.31}) are satisf\/ied, the solution of the two equations (\ref{San2}) is given by
\begin{gather*}
\theta=\int_{u_{0,0}^{\prime}=\alpha}^{u_{0,0}^{\prime}=u_{0,0}}\frac{W_{(u_{-1,0})}  [\mathcal C;\mathcal D  ]} {\mathcal C_{,u_{-1,0}}} (u_{0,0}^{\prime},u_{0,1} )du_{0,0}^{\prime}-\int_{u_{0,1}^{\prime}=\beta}^{u_{0,1}^{\prime}=u_{0,1}}\frac{\mathcal D_{,u_{-1,0}}}{\mathcal C_{,u_{-1,0}}} (\alpha,u_{0,1}^{\prime} )du_{0,1}^{\prime}+\gamma,
\end{gather*}
where $\alpha$ and $\beta$ are some f\/ixed values  of the variables $u_{0,0}$ and $u_{0,1}$ at which the integrals are well def\/ined, while $\gamma$ is an arbitrary integration constant.

\section{Examples}\label{section5}

Here we consider the linearizability conditions in the case of some interesting examples.

\subsection{Liouville equation}

Let us consider the discrete Liouville equation \cite{s11}
\begin{gather}\label{5.8a}
u_{1,1}=\frac{(u_{1,0}-1)(u_{0,1}-1)}{u_{0,0}}\doteq F(u_{0,0}, u_{0,1}, u_{1,0}).
\end{gather}
In \cite{s11} it was shown that the transformation
\[ 
u_{0,0}=\frac{\tilde u_{1,0}\tilde u_{0,1}}{(\tilde u_{1,0}-\tilde u_{0,0})(\tilde u_{0,1}-\tilde u_{0,0})},
\]
 maps solutions of the linear equation
\[
\tilde u_{0,0}-\tilde u_{1,0}-\tilde u_{0,1}+\tilde u_{1,1}=0,
\]
into solutions of~(\ref{5.8a}). This transformation is not of the form considered here as depends on three points. However this example might also be linearized by the transformations considered in the previous sections.

We can try to linearize this discrete Liouville equation by a point transformation. The ne\-ces\-sary conditions (\ref{3.13b}), (\ref{3.13c}) are identically satisf\/ied and the linearizing point transformation, obtained by integrating  (\ref{3.11a}), is given by
\[
f (u_{0,0} )=A [B+\log (u_{0,0}-1 ) ],
\]
where $A\not=0$ and $B$ are arbitrary constants. One can easily see that it does not exist any value of the constants $B$, $a\not=0$, $b\not=0$ and $c\not=0$ for which the function $f (u_{0,0} )$ satisf\/ies (\ref{3.7a}) identically modulo~(\ref{5.8a}) (the multiplicative constant~$A$ is inessential as it can be always rescaled away). Hence the Liouville equation~(\ref{5.8a}) \emph{cannot be linearized by a~point transformation}.

We can try to linearize by a two points transformation of the form (\ref{1.3c}). As $\mathbb F= \mathbb K=0$ identically, we are in the fourth case. Moreover the two linearizabilty conditions (\ref{cond1}), (\ref{cond2}) are identically satisf\/ied and the overdetermined system of dif\/ferential equations (\ref{3.8amd}) reads
\begin{gather}\label{Ali}
 g_{u_{0,0},u_{0,0}}-\frac{1-u_{0,1}}{u_{0,0}}g_{u_{0,0},u_{0,1}}+\frac{1}{u_{0,0}}g_{u_{0,0}}=0,\\\nonumber
 g_{u_{0,1},u_{0,1}}-\frac{u_{0,0}}{1-u_{0,1}}g_{u_{0,0},u_{0,1}}-\frac{1}{1-u_{0,1}}g_{u_{0,1}}=0,
\end{gather}
whose solution is given by
\begin{gather}\label{Ali1}
g\left(u_{0,0},u_{0,1}\right)=\theta\left(\xi\right)+C\log\left(u_{0,0}\right)+D,\qquad \xi=\frac{u_{0,1}-1}{u_{0,0}},
\end{gather}
where $C$ and $D$ are arbitrary constants and $\theta\not=0$ is an arbitrary function of its arguments. As one can see, the system (\ref{Ali}) does not  specify the  two points transformation. To def\/ine it we need to introduce (\ref{Ali1}) into (\ref{3.6aa}), (\ref{3.6aaab}). In this way  we get a system of two f\/irst order dif\/ferential-dif\/ference equations involving $\theta$ and $T_1 \theta$. From them we can extract a f\/irst order ordinary dif\/ferential equation for $\theta$ which depends on~$\xi$ and~$u_{1,0}$. As a consequence this equation splits into an overdetermined system of two f\/irst order ordinary dif\/ferential equations for $\theta(\xi)$, whose solution is given by
\[
\theta=C\log (\xi+1 )+\alpha, \qquad b=-1, \qquad a=-c,
\]
where $\alpha$ is an arbitrary constant and $C\not=0$. Hence, after a reparametrization of $D$,
\begin{gather} \label{nn1}
g (u_{0,0},u_{0,1} )=C [\log (u_{0,0}+u_{0,1}-1 )+D ].
\end{gather}
 A necessary condition for (\ref{nn1}) to be a linearizing transformation, is that (\ref{m1.3a}) be identically satisf\/ied modulo~(\ref{5.8a}). It is easy to show that it is not possible to f\/ind a value of $D$ and $C\not=0$ such that   this condition is satisf\/ied. In conclusion the equation~(\ref{5.8a}) \emph{cannot be linearized by a~two points transformation}.

 If we consider the linearization through a Hopf--Cole transformation, we are in the case when $\mathbb F=\mathbb K=0$ and the linearizability conditions $\mathcal R_{,u_{1,0}}=\mathcal U_{,u_{1,0}}=0$ are also satisf\/ied. The  equations for the functions  $\psi$ and $\phi$ read
\begin{gather*}
\phi_{,u_{0,1}}+\frac{u_{0,0}}{u_{0,1}-1}\phi_{,u_{0,0}}+\frac{1}{u_{0,1}-1}=0, \qquad
\psi_{,u_{0,-1}}+\frac{u_{0,0}-1}{u_{0,-1}}\psi_{,u_{0,0}}+\frac{1}{u_{0,0}}=0,
\end{gather*}
and their solution imply
\begin{gather}\label{San12}
h+1/b=\sigma (\tilde\xi )\rho (u_{0,0} ), \qquad\frac{h+1/b}{h}=\kappa (\tilde\xi )\tau (u_{0,1} ), \qquad\tilde\xi\doteq\frac{u_{0,1}-1}{u_{0,0}},
\end{gather}
where $\sigma$, $\rho$, $\kappa$ and $\tau$ are arbitrary nonzero functions of their argument. The function $\theta (u_{0,0},u_{0,1} )$ def\/ined in (\ref{San3}), is specif\/ied by the conditions $\mathcal C_{,u_{-1,0}}=\mathbb F=0$. The necessary conditions $\mathcal D_{,u_{-1,0}}=\tilde {\mathcal T}_{,u_{1,0}}=0$ are satisf\/ied and the two equations (\ref{San1}), (\ref{San5}) respectively read
\[
\theta_{,u_{0,0}}+\frac{u_{0,1}}{u_{0,0}-1}\theta_{,u_{0,1}}=0, \qquad\theta_{,u_{0,1}}+\frac{u_{0,0}}{u_{0,1}-1}\theta_{,u_{0,0}}=0.
\]
The only admissible solution of this system is  a  constant. The solution of the overdetermined system of the two functional equations (\ref{San12}) and of the Hopf-like partial dif\/ferential equation~(\ref{San4}), after a reparametrization of the constant $\theta$, is given by
\[
h=-\frac{\gamma (u_{0,1}-1 )+\delta}{b\delta+\tilde\theta u_{0,0}},
\]
where $\gamma\not=0$, $\delta$ and $\tilde\theta\not=0$ are arbitrary constants. A necessary condition to obtain a linearizing transformation is that~(\ref{3.6}) be identically satisf\/ied modulo (\ref{5.8a}). No nonzero value of $a$, $b$, $c$, $\gamma$, $\tilde\theta$ and $\delta$ can satisfy this condition, hence (\ref{5.8a}) \emph{cannot be linearized by a Hopf--Cole transformation too}.

\subsection{Second Liouville equation}

Let us consider the following version of the discrete Liouville equation
\begin{gather}\label{5.8b}
w_{1,1}=\frac{w_{1,0}w_{0,1}-1}{w_{0,0}}\doteq F (w_{0,0}, w_{0,1}, w_{1,0} ).
\end{gather}
As shown in \cite{s11},  the noninvertible transformation $u_{0,0}=w_{1,0}w_{0,1}$ maps solutions of (\ref{5.8b}) into solutions of (\ref{5.8a}).

Let us look for a linearizing point  transformation. The necessary conditions (\ref{3.13c}) are identically satisf\/ied while condition (\ref{3.13b}) reads $-1/u_{0,0}^3=0$. Hence we can conclude that (\ref{5.8b}) \emph{cannot be linearized by a point transformation}.

Let us look for a linearizing two points transformation. We are in the subcase~(1)  as $\mathbb F=-\mathbb K=-1/u_{0,0}^{3}\not=0$ and $\mathcal P=-1/u_{0,0}\not=0$. Moreover we have $\mathcal M=-1/u_{0,0}$, \mbox{$\mathcal N=\mathcal Q=0$}. The integrability conditions (\ref{3.8ag}), (\ref{3.8af}), (\ref{3.8agcc}), (\ref{3.8afc}) are identically satisf\/ied while the conditions (\ref{Maro}) cannot be satisf\/ied. As a consequence (\ref{5.8b}) \emph{cannot be linearized by a~two points transformation}.

If we consider the linearization through a Hopf--Cole transformation, we are in the case where $\mathbb F\not=0$, $\mathbb K \not=0$ and each of the two equations (\ref{3.14}), (\ref{3.14a}) splits into  two equations
\begin{gather*}
 \phi_{,u_{0,0}} = \mathcal S (u_{0,0},u_{1,0},u_{0,1} )\doteq\frac{\mathcal R_{,u_{1,0}}F_{,u_{0,0}}^2}{\mathbb F}, \qquad\phi_{,u_{0,1}}=\mathcal R+\frac{F_{,u_{0,1}}}{F_{,u_{0,0}}}\mathcal S,\\
 \psi_{,u_{0,0}} = \mathcal V (u_{0,-1},u_{0,0},u_{1,0} )\doteq\frac{\mathcal U_{,u_{1,0}}K_{,u_{0,0}}^2}{\mathbb K}, \qquad\psi_{,u_{0,-1}}=\mathcal U+\frac{K_{,u_{0,-1}}}{K_{,u_{0,0}}}\mathcal V,
\end{gather*}
which respectively read
\[
\phi_{,u_{0,0}}+\frac{1}{u_{0,0}}=0, \qquad \phi_{,u_{0,1}}=0, \qquad\psi_{,u_{0,0}}+\frac{1}{u_{0,0}}=0, \qquad \psi F_{,u_{0,-1}}=0.
\]
The necessary conditions $(\mathcal S_{,u_{1,0}}, \mathcal V_{,u_{1,0}})=(0,0)$ are respected and the solutions of the two equations respectively imply
\[
 h+\frac{1}{b}=e^{\alpha\frac{u_{0,1}}{u_{0,0}}}\rho (u_{0,0} ), \qquad\frac{h+1/b}{h}=e^{\beta\frac{u_{0,0}}{u_{0,1}}}\tau (u_{0,1} ),
\]
where $\alpha$ and $\beta$ are two arbitrary nonzero constants and $\rho$ and $\tau$ are arbitrary nonzero functions of their arguments. It is not dif\/f\/icult to see that no $\alpha$, $\beta$, $\theta$ and $\tau$ exist, giving a nontrivial, e.g.\ nonconstant, solution for $h$. Hence (\ref{5.8b}) \emph{cannot be linearized by a Hopf--Cole transformation too}.

\subsection[$\mathcal Q_+$ equation  linearizable upto  $5^{\rm th}$ order by a multiple scale expansion \cite{hls}]{$\boldsymbol{\mathcal Q_+}$ equation  linearizable upto  $\boldsymbol{5^{\rm th}}$ order by a multiple scale expansion \cite{hls}}

Let us consider the equation
\begin{gather}
\zeta u_{0,0}u_{1,0}u_{0,1}u_{1,1}+a_{1} (u_{0,0}+u_{1,1} )+a_{2} (u_{1,0}+u_{0,1} )+\gamma_{1}u_{0,0}u_{1,1} \nonumber \\
\qquad {} +\frac{a_{2}\gamma_{1}}{a_{1}}u_{1,0}u_{0,1}+\frac{(a_{1}+a_{2} )\gamma_{1}}{2a_{1}} (u_{0,0}u_{0,1}+u_{1,0}u_{1,1}+u_{0,0}u_{1,0}+u_{0,1}u_{1,1} ) \nonumber\\
\qquad {} +\frac{(a_{1}+2a_{2})\gamma_{1}^{2}}{4a_{1}^{2}}(u_{0,0}+u_{1,1})u_{1,0}u_{0,1}+\frac{(2a_{1}+a_{2})\gamma_{1}^{2}}{4a_{1}^{2}}(u_{1,0}+u_{0,1})u_{0,0}u_{1,1},
\label{Orma}
\end{gather}
where $a_{1}$, $a_{2}$, $\gamma_{1}$, $\zeta$ are arbitrary real parameters with $|a_{1}|\not=|a_{2}|$ and $a_{j}\not=0$, $j=1, 2$. In~\cite{hls} it has been shown that this equation passes a linearizability test based on multiscale analysis up to f\/ifth order in the perturbation parameter for small~$u$.

Let us search for the possibility to linearize (\ref{Orma}) by a point transformation. Of the necessary conditions (\ref{3.13c}),  $\mathcal H^{(0)}=0$ is automatically satisf\/ied while  $\mathcal K^{(0)}=0$ and (\ref{3.13b}) can be satisf\/ied  if and only if
\begin{gather}\label{Amor}
\zeta=\frac{\left(a_{1}+a_{2}\right)\gamma_{1}^3}{4a_{1}^{3}}.
\end{gather}
In this case the linearizing point transformation obtained integrating (\ref{3.11a}) is given by
\begin{gather} \label{nn2}
f (u_{0,0} )=A\left[\frac{1}{2a_{1}+\gamma_{1}u_{0,0}}+B\right],
\end{gather}
where $A\not=0$ and $B$ are constants. Inserting $f (u_{0,0} )$ into (\ref{3.7a}), one f\/inds that this relation is identically satisf\/ied modulo (\ref{Orma}), (\ref{Amor}) when $c=ba_{1}/a_{2}$. Finally, inserting $f (u_{0,0} )$ and~$c$ into~(\ref{3.4a}), it is straightforward to see that this relation results identically satisf\/ied modulo (\ref{Orma}), (\ref{Amor}) when $B=-1/(2a_{1})$, $c=1$ and $a=b=a_2/a_1$. Equation~(\ref{nn2}) is the linearizing point transformation when $B=-1/(2 a_1)$.

Let us consider the case $\zeta\not= (a_{1}+a_{2} )\gamma_{1}^3/(4a_{1}^{3})$. If we search for a linearizing two points transformation, as $a_{j}\not=0$, $j=1, 2$, we are always in the subcase~$(a)$. Then the necessary condition~(\ref{3.8ag}) cannot be satisf\/ied. So, if the condition~(\ref{Amor}) is not satisf\/ied, (\ref{Orma}) \emph{cannot be linearized by a two points transformation}.

We can try to linearize (\ref{Orma}) by a Hopf--Cole transformation. We are in the case where $\mathcal F \not=0$ and the equation (\ref{3.14}) splits into two equations. As the necessary condition $\mathcal S_{,u_{1,0}}=0$ cannot be satisf\/ied, (\ref{Orma}) \emph{cannot be linearized by a Hopf--Cole transformation}. We can conclude that, if the condition (\ref{Amor}) is not satisf\/ied, (\ref{Orma}) \emph{cannot be linearized by neither a point, nor contact or Hopf--Cole transformation}.

\subsection{Hietarinta equation}

Let us consider the Hietarinta equation \cite{h04,jrg}
\begin{gather}\label{Rosa}
\frac{u_{0,0}+e_{2}} {u_{0,0}+e_{1}} \frac{u_{1,1}+o_{2}} {u_{1,1}+o_{1}}=\frac{u_{1,0}+e_{2}} {u_{1,0}+o_{1}} \frac{u_{0,1}+o_{2}} {u_{0,1}+e_{1}},
\end{gather}
where $e_{j}$ and $o_{j}$, $j=1, 2$ are arbitrary parameters.

\subsubsection{Linearizing one point transformation}

The necessary conditions (\ref{3.13b}), (\ref{3.13c}) of linearizability  through a point transformation are identically satisf\/ied and the integration of equation~(\ref{3.11a}) gives
\[
f (u_{0,0} )=A \left[\log\left (\frac{u_{0,0}+o_{1}}{u_{0,0}+o_{2}} \right)+B \right],
\]
where $A\not=0$ and $B$ are arbitrary constants. One can easily see that no values of the constants~$B$, $a\not=0$, $b\not=0$ and $c\not=0$ exist for which the function $f (u_{0,0})$ satisf\/ies (\ref{3.7a}) identically modulo the Hietarinta equation. As a consequence~(\ref{Rosa}) \emph{cannot be linearized by a point transformation}.

\subsubsection{Linearizing two points transformation}

As $\mathbb F=\mathbb K=0,$  we are in the  case (4). Moreover the two linearizability conditions (\ref{cond1}), (\ref{cond2}) are identically satisf\/ied and the overdetermined system of dif\/ferential equations (\ref{3.8amd}) reads
\begin{gather} \label{Ale}
g_{,u_{0,0},u_{0,0}}\!+\frac{(e_{2}-e_{1})(u_{0,1}+o_{1})(u_{0,1}+o_{2})}{(o_{2}-o_{1})(u_{0,0}+e_{1})(u_{0,0}+e_{2})}g_{,u_{0,0},u_{0,1}}\!
+\frac{2u_{0,0}+e_{1}+e_{2}}{(u_{0,0}+e_{1})(u_{0,0}+e_{2})}g_{,u_{0,0}}\!=0,\!\!\!\\ \nonumber
g_{,u_{0,1},u_{0,1}}+\frac{(o_{2}-o_{1})(u_{0,0}+e_{1})(u_{0,0}+e_{2})}{(e_{2}-e_{1})(u_{0,1}+o_{1})(u_{0,1}+o_{2})}g_{,u_{0,0},u_{0,1}}
+\frac{2u_{0,1}+o_{1}+o_{2}}{(u_{0,1}+o_{1})(u_{0,1}+o_{2})}g_{0,0,u_{0,1}}=0.
\end{gather}
The solution of (\ref{Ale}) is given by
\begin{gather}\label{Ale1}
g (u_{0,0},u_{0,1} )=\theta(\xi)+A\log\left(\frac{u_{0,1}+o_{2}}{u_{0,1}+o_{1}}\right)+B, \qquad \xi=\frac{ (u_{0,0}+e_{2} ) (u_{0,1}+o_{1})}{(u_{0,0}+e_{1})(u_{0,1}+o_{2})},
\end{gather}
where $A$ and $B$ are arbitrary constants and $\theta\not=0$ is an arbitrary function of its argument. As one can see, the system (\ref{Ale}) is not suf\/f\/icient to specify the eventual two points transformation. We need to introduce (\ref{Ale1}) into (\ref{3.6aa}), (\ref{3.6aaab}). In this way we get a system of f\/irst order dif\/ferential equations involving $\theta$ and $T_1 \theta$. From them we can extract a f\/irst order ordinary dif\/ferential equation for $\theta$ which depends on $\xi$, $u_{0,0}$ and $u_{0,1}$. As a consequence this equation splits into an overdetermined system of four ordinary dif\/ferential equations for $\theta(\xi)$. This system has no solution for generic $e_{j}$, $o_{j}$, $j=1, 2$. As a consequence the Hietarinta equation \emph{cannot be linearized by a two points transformation}.

\subsubsection[Linearizing one point Hopf-Cole transformation]{Linearizing one point Hopf--Cole transformation}

We are in the case when (\ref{3.13b}), (\ref{3.13c}) are satisf\/ied, and the integration of (\ref{Reghini}) gives
\[
 h (u_{0,0})=\frac{1}{b}\left[A\left(\frac{u_{0,0}+e_{1}}{u_{0,0}+o_{2}}\right)^B-1\right],
\]
where $A\not=0$ and $B\not=0$ are arbitrary integration constants. We have that (\ref{Amaryllis}) can be identically satisf\/ied modulo the Hietarinta equation if and only if
\[
A=\left(1-\frac{ab}{c}\right)\left(\frac{o_{2}-o_{1}}{e_{1}-o_{1}}\right),\qquad  B=1.
\]
Finally (\ref{3.7}) is identically satisf\/ied modulo the Hietarinta equation if and only if
\[
A=-\frac{o_{2}-e_{2}}{e_{2}-e_{1}}, \qquad a=\frac{c(e_{2}-o_{1})(o_{2}-e_{1})}{b(e_{2}-e_{1})(o_{2}-o_{1})},
\]
so that
\[
\tilde u_{0,1}=-\frac{1}{b}\frac{(o_{2}-e_{1})(u_{0,0}+e_{2})}{(e_{2}-e_{1})(u_{0,0}+o_{2})}\tilde u_{0,0}.
\]
Through the gauge transformation $\tilde u_{0,0}\doteq (-b/c )^{n} (-b )^{-m}w_{0,0}$ we get a simplif\/ied linearizing transformation
\begin{gather}
w_{0,1} = \frac{(e_{1}-o_{2})(u_{0,0}+e_{2})}{(e_{1}-e_{2})(u_{0,0}+o_{2})}w_{0,0},\label{Rumon3}\\
w_{0,0}+aw_{1,0}-w_{0,1}+w_{1,1}=0,\qquad a=-\frac{(o_{1}-e_{2})(e_{1}-o_{2})}{(e_{1}-e_{2})(o_{1}-o_{2})}\label{Rumon4}.
\end{gather}
Is is moreover straightforward to demonstrate that if (\ref{Rumon3}), (\ref{Rumon4}) are satisf\/ied, then also the Hietarinta equation is satisf\/ied.

\subsubsection[Linearizing two point Hopf-Cole transformation \cite{jrg}]{Linearizing two point Hopf--Cole transformation \cite{jrg}}

We are in the case def\/ined by the conditions $\mathcal C_{,u_{-1,0}}=\mathcal D_{,u_{-1,0}}=\mathcal T_{,u_{1,0}}=\mathbb F=0$ and thus the linearizing function is def\/ined by the equations (\ref{San1}), (\ref{San5}) which read
\begin{gather*}
\theta_{,u_{0,0}}+\frac{(e_{2}-o_{1})(u_{0,1}+o_{1})(u_{0,1}+o_{2})}{(o_{2}-o_{1})(u_{0,0}+o_{1})(u_{0,0}+e_{2})}\theta_{,u_{0,1}}
=2\frac{u_{0,0}(o_{2}-o_{1})-u_{0,1}(e_{2}-o_{1})+o_{1}(o_{2}-e_{2})}{(o_{2}-o_{1})(u_{0,0}+o_{1})(u_{0,0}+e_{2})},\\\nonumber
\theta_{,u_{0,1}}+\frac{(o_{2}-e_{1})(u_{0,0}+e_{1})(u_{0,0}+e_{2})}{(e_{2}-e_{1})(u_{0,1}+e_{1})(u_{0,1}+o_{2})}\theta_{,u_{0,0}}
=2\frac{u_{0,0}(o_{2}-e_{1})-u_{0,1}(e_{2}-e_{1})+e_{1}(o_{2}-e_{2})}{(e_{2}-e_{1})(u_{0,1}+e_{1})(u_{0,1}+o_{2})},
\end{gather*}
whose solution is given by
\[
\theta (u_{0,0},u_{0,1} )=\log\left[\left(\frac{u_{0,0}+e_{2}}{u_{0,1}+o_{2}}\right)^2\right]+\alpha,
\]
where $\alpha$ is an arbitrary integration constant. Then we can solve the Hopf-like equation (\ref{San4}) by separation of variables,   $h=A (u_{0,0} )B (u_{0,1})$, obtaining
\begin{gather}\label{San10}
h(u_{0,0},u_{0,1})=\frac{e^{\alpha}(u_{0,0}+e_{2})}{\delta(u_{0,0}+e_{2})+\beta} \frac{\gamma(u_{0,1}+o_{2})-\beta}{u_{0,1}+o_{2}},
\end{gather}
where $\beta$, $\gamma$ and $\delta$ are arbitrary integration constants. A necessary condition to obtain the linearization is that~(\ref{3.6}) be identically satisf\/ied for all $u_{0,0}$, $u_{1,0}$, $u_{0,1}$, $u_{0,2}$, $e_{j}$, $o_{j}$, $j=1, 2$ modulo the Hietarinta equation, from which we get
\begin{gather*}
e^{\alpha}=\frac{a\beta(o_{2}-o_{1})}{c(e_{2}-o_{1})[\beta+\gamma(e_{2}-o_{2})]}, \qquad\delta=-\frac{\beta\gamma}{\beta+\gamma(e_{2}-o_{2})}, \qquad b=\frac{c(o_{2}-e_{1})(e_{2}-o_{1})}{a(e_{2}-e_{1})(o_{2}-o_{1})}.
\end{gather*}
When we insert the obtained values of $e^{\alpha}$, $\delta$ and $b$ into the transformation~(\ref{San10}), the two equations for $\psi$ and $\phi$ are identically satisf\/ied. As~$h$  depends on $u_{0,0}$ and $u_{0,1}$, it is necessary that $\beta\not=0$. By redef\/ining $\gamma\doteq\beta\epsilon$ we can eliminate the parameter $\beta$ from the transformation. The transformation as well as  the coef\/f\/icient $b$ of the linear equation  so far obtained depend in a multiplicative way from the ratio~$a/c$. Hence, performing the transformation $\tilde u_{n,m}\doteq\chi ^{m} v_{n,m}$, where $\chi$ is a constant, the linear equation~(\ref{1.3a}) and the Hopf--Cole transformation~(\ref{1.3c}) respectively read
\begin{gather*}
 v_{n,m}+av_{n+1,m}+b\chi v_{n,m+1}+c\chi v_{n+1,m+1}=0,\\
  v_{n,m+1}=\frac{1}{\chi}h (u_{n,m},u_{n,m+1} )v_{n,m},
\end{gather*}
and choosing $\chi=-a/c$ we can remove the ratio $a/c$ from the expressions of $h$ and $b$. In other words we can always choose a ``gauge'' for the linearizing transformation in which $a=-c$. In conclusion we have
\begin{gather}\label{Italia}
h=\frac{(o_{2}-o_{1})(u_{0,0}+e_{2})[1-\epsilon(u_{0,1}+o_{2})]}{(e_{2}-o_{1})[1-\epsilon(u_{0,0}+o_{2})](u_{0,1}+o_{2})}, \qquad b=-\frac{(o_{2}-e_{1})(e_{2}-o_{1})}{(e_{2}-e_{1})(o_{2}-o_{1})}.
\end{gather}

\subsubsection[Inverse two point Hopf-Cole transformation]{Inverse two point Hopf--Cole transformation}

To be able to solve the Hietarinta equation explicitly we  look here for the inverse formula, i.e.\ the formula which provide the solution of the Hietarinta equation in terms of those of the linear equation~(\ref{1.3a}).

 Given a solution of the Hietarinta equation, let us perform a Hopf--Cole transformation~(\ref{1.3d}) with~$h$ given in~(\ref{Italia}). Extracting from the Hopf--Cole transformation $u_{0,1}$ as a function of $u_{0,0}$ and $\eta_{0,0}\doteq\tilde u_{0,1}/\tilde u_{0,0}$  and inserting it,  together with its consequences, in the Hietarinta equation, if $\epsilon\not=\frac{1}{(o_{2}-o_{1})}$, we obtain $u_{1,0}$ as a function of $u_{0,0}$, $\eta_{0,0}$ and $\eta_{1,0}$. The compatibility between the functions $u_{0,1}(u_{0,0},\eta_{0,0})$ and $u_{1,0}(u_{0,0},\eta_{0,0},\eta_{1,0})$ gives a second degree polynomial equation in~$u_{0,0}$ which must be satisf\/ied for all $u_{0,0}$. Taking into account that the ratio $\eta_{0,0}$ cannot in general be a constant when $u_{0,0}$ satisf\/ies the Hietarinta equation, we have that, if $\epsilon\not=\frac{1}{(o_{2}-o_{1})}$, $\tilde u_{0,0}$ will satisfy the linear and in general nonautonomous equation
\begin{gather}\label{Latium}
(1-\alpha_{n})\tilde u_{0,0}-c\tilde u_{1,0}+b(1-\alpha_{n})\tilde u_{0,1}+c\tilde u_{1,1}=0, \qquad c=\frac{(e_{1}-o_{1})(e_{2}-o_{1})}{(e_{2}-e_{1})(o_{2}-o_{1})},
\end{gather}
where $\alpha_{n}$ is an $n$-dependent integration function depending on the initial values $u_{n,0}$ and $\tilde u_{n,0}$ given by
\begin{gather}\label{Urbs}
\alpha_{n}=1-\frac{(e_{1}-o_{1})[1-\epsilon(u_{n,0}+o_{2})](u_{n+1,0}+o_{1})\tilde u_{n+1,0}}{(o_{2}-o_{1})(u_{n,0}+e_{1})[1-\epsilon(u_{n+1,0}+o_{2})]\tilde u_{n,0}}.
\end{gather}
When $\alpha_{n}\not=1$, performing the ``gauge'' transformation $\tilde u_{n,m}\doteq\tau_{n}v_{n,m}$, with $\tau_{n+1}=(1+\alpha_{n})\tau_{n}$, the Hopf--Cole transformation is invariant while the function $v_{n,m}$ will satisfy the linear autonomous equation
\begin{gather}\label{Atli}
v_{0,0}-cv_{1,0}+bv_{0,1}+cv_{1,1}=0.
\end{gather}

When $\epsilon=\frac{1}{(o_{2}-o_{1})}$, the function $h$ becomes
\begin{gather}\label{San6}
h=\frac{(o_{2}-o_{1})(u_{0,0}+e_{2})(u_{0,1}+o_{1})}{(e_{2}-o_{1})(u_{0,0}+e_{1})(u_{0,1}+o_{2})}.
\end{gather}
Inserting the corresponding Hopf--Cole transformation into the Hietarinta equation we get $u_{0,0}$ in terms of $\eta_{0,0}$ and $\eta_{1,0}$
\begin{gather}\label{AmorAle}
u_{0,0}=-\frac{e_{2}(e_{1}-o_{1})(o_{2}-o_{1})+o_{1}(e_{2}-o_{1})(o_{2}-e_{1})\eta_{0,0}
-e_{1}(e_{2}-o_{1})(o_{2}-o_{1})\eta_{1,0}}{(e_{1}-o_{1})(o_{2}-o_{1})
+(e_{2}-o_{1})(o_{2}-e_{1})\eta_{0,0}-(e_{2}-o_{1})(o_{2}-o_{1})\eta_{1,0}}
\end{gather}
and its insertion with its consequences into $u_{0,1}(u_{0,0},\eta_{0,0})$ implies for the function $\tilde u_{0,0}$ the same evolution (\ref{Latium}) with $\alpha_{n}$ given by the limit as $\epsilon\rightarrow\frac{1}{(o_{2}-o_{1})}$ of (\ref{Urbs}) and hence (\ref{Atli}) for $v_{0,0}$, just as was shown in~\cite{jrg}. As a consequence $u_{0,0}$, given by (\ref{AmorAle}), becomes
\begin{gather}\label{Orbs}
u_{0,0}=-\frac{e_{1}(o_{2}-o_{1})v_{0,0}-o_{1}(e_{1}-o_{1})v_{1,0}}{(o_{2}-o_{1})v_{0,0}-(e_{1}-o_{1})v_{1,0}}.
\end{gather}
Conversely, if $v_{0,0}$ satisf\/ies (\ref{Atli}), then it is possible to demonstrate that $u_{0,0}$ given by (\ref{Orbs}) satisf\/ies the Hietarinta equation.

 In the relations (\ref{AmorAle}), ({\ref{Orbs}}) the role of the variables $u_{n,m}$ and $v_{n,m}$ appears inverted with respect to those given by the Hopf--Cole transformation (\ref{1.3d}), (\ref{1.3e}).

 The inverted transformation, consequently, appears only when $\epsilon\!=\!\frac{1}{(o_{2}{-}o_{1})}$. The rela\-tion~(\ref{Orbs}) is the inverse in the space of the solutions of the Hietarinta equation of the relation (\ref{1.3e}) with the function $k$ def\/ined in (\ref{1.3f}) corresponding to an $h$ as given in (\ref{San6}) and restricting the result to the space of the solutions of the Hietarinta equation, that is
\begin{gather}\label{Juno}
\frac{v_{1,0}}{v_{0,0}}=k=\frac{(o_{2}-o_{1})(u_{0,0}+e_{1})}{(e_{1}-o_{1})(u_{0,0}+o_{1})}.
\end{gather}
Finally, inserting the initial value at $m=0$ of (\ref{Juno}) into (\ref{Urbs}), we get $\alpha_{n}=0$. By the transformation $v_{0,0}\doteq c^{-n} w_{0,0}$, we can simplify  further (\ref{Atli}) which together with (\ref{Juno}) becomes
\begin{gather}
\tilde w_{0,0}-\tilde w_{1,0}+b\tilde w_{0,1}+\tilde w_{1,1}=0, \qquad
 \tilde w_{1,0}=\frac{(e_{2}-o_{1})(u_{0,0}+e_{1})}{(e_{2}-e_{1})(u_{0,0}+o_{1})}\tilde w_{0,0}.\label{Ovidio}
\end{gather}
The second relation in (\ref{Ovidio}) represents another linearizing one point Hopf--Cole transformation  $\tilde w_{1,0}=\tilde h(u_{0,0})\tilde w_{0,0}$.

\subsubsection{Solution of the Hietarinta equation}

The general integral of the Hietarinta equation is obtained inserting in the inverse of the second relation~(\ref{Ovidio}) the solution of the initial-boundary value
problem for the linear equation given in~(\ref{Ovidio}).  This
solution is given by
\begin{itemize}\itemsep=0pt
\item if $b=-1$ by $\tilde w_{n,m}=\tilde w_{n,0}+\tilde w_{0,m}-\tilde w_{0,0}$,
\item if $b\not=-1$ by $\tilde w_{n,m}=\frac{1}{2\pi\ri}\oint_{C}\zeta_{m} (z )z^{n-1}dz$, where
\[
 (1-z )\zeta_{m}+ (b+z )\zeta_{m+1}=z (\tilde w_{0,m+1}-\tilde w_{0,m} ), \qquad \zeta_{0} (z )=\sum_{j=0}^{+\infty}\tilde w_{j,0}z^{-j},
\]
and $C$ represents a counterclockwise circumference in the complex $z-$plane, centered in $z=0$ and lying inside the region of convergence of the series $\sum\limits_{j=0}^{+\infty}\tilde w_{j,m}z^{-j}$.
\end{itemize}

 For example, when $b\not=-1$, an explicit solution in the plane $n\geq 0$ of the initial-boundary value problem characterized by $\tilde w_{0,m}=1$, $\tilde w_{n,0}=2^{-n}$, $n\geq 0$ ($\tilde w_{n,0}=0$, $n<0$), so that $\zeta_{0}=2z/(2z-1)$, $|z|>1/2$, is given, if $b=-\frac{1}{2}$, by
\begin{gather*}
  \tilde w_{n,m} = \sum_{\rho=\max\{n,m\}}^{n+m}\begin{pmatrix}
m\\
\rho-n\end{pmatrix} \begin{pmatrix}
\rho\\
m\end{pmatrix}\frac{(-1)^{n+m-\rho}}{2^{\rho-m}}, \qquad m\geq 0,\\
 \tilde w_{n,m} = \sum_{\rho=\max\{n,|m|-1\}}^{n+|m|-1}\begin{pmatrix}
|m|-1\\
\rho-n\end{pmatrix}\begin{pmatrix}
\rho\\
|m|-1\end{pmatrix}\left(-\frac{1}{2}\right)^{n+|m|-\rho-1}, \qquad m\leq -1,
\end{gather*}
and, if $b\not=-\frac{1}{2}$, by
\begin{gather*} \tilde w_{n,m} = \frac{1}{2^{n}(-1-2b)^{m}}-\sum_{\kappa=0}^{m-1}\sum_{\rho=\max\{n,\kappa\}}^{n+m}\begin{pmatrix}
m\\
\rho-n\end{pmatrix}\begin{pmatrix}
\rho\\
\kappa\end{pmatrix}\frac{(-1)^{n+m-\kappa}b^{\rho-\kappa}}{(b+1/2)^{m-\kappa}},\qquad m\geq1,\\
  \tilde w_{n,m} = \frac{(-1-2b)^{|m|}}{2^{n}}-\sum_{\kappa=0}^{|m|-1}\sum_{\rho=\max\{n,\kappa\}}^{n+|m|}\begin{pmatrix}
|m|\\
\rho-n\end{pmatrix}\begin{pmatrix}
\rho\\
\kappa\end{pmatrix}(-2)^{|m|-\kappa}b^{n+|m|-\rho},\qquad m\leq-1.
\end{gather*}

\subsection{A nonlinear quad-graph equation linearizable\\ by two point transformation}

 As a further example let us consider the simple multilinear equation
\begin{gather} \label{n1}
u_{0,0} + \alpha u_{1,0} + \epsilon [ u_{0,0} u_{0,1} + \alpha u_{1,0} u_{1,1}] = 0.
\end{gather}
As in the case of the discrete Liouville equation the necessary conditions for the linearizability by point transformations are identically satisf\/ied and  the linearizing point transformation, obtained by integrating  (\ref{3.11a}), is given by
\[ 
f (u_{0,0} )=A\left[B+\log\left(u_{0,0}+\frac{1}{\epsilon}\right)\right],
\]
where $A\not=0$ and $B$ are arbitrary constants. Inserting $f\left(u_{0,0}\right)$ into (\ref{3.7a}), one can easily see that this relation can be identically satisf\/ied modulo (\ref{n1}) only if $c=-b$. Dif\/ferentiating (\ref{3.4a}) with respect to $u_{1,0}$ and inserting $f\left(u_{0,0}\right)$ and $c=-b$ into the resulting relation, we have that no values of the constants~$B$, $a\not=0$, $b\not=0$ and $c\not=0$ exist for which  this can be satisf\/ied identically modulo~(\ref{n1}). Hence this equation (\ref{n1}) \emph{cannot be linearized by a point transformation}.

We can try to linearize it by a two points transformation of the form (\ref{1.3c}). As $\mathbb F= \mathbb K=0$ identically, we are in the fourth case. Moreover the two linearizabilty conditions (\ref{cond1}), (\ref{cond2}) are identically satisf\/ied and the overdetermined system of dif\/ferential equations (\ref{3.8amd}), (\ref{3.8ame}) reads
\begin{gather*}
g_{u_{0,0},u_{0,0}}-\frac{1+\epsilon u_{0,1}}{\epsilon u_{0,0}}g_{u_{0,0},u_{0,1}}+\frac{1}{u_{0,0}}g_{u_{0,0}}=0,\\ \nonumber
g_{u_{0,1},u_{0,1}}-\frac{\epsilon u_{0,0}}{1+u\epsilon_{0,1}}g_{u_{0,0},u_{0,1}}+\frac{\epsilon}{1+\epsilon u_{0,1}}g_{u_{0,1}}=0,
\end{gather*}
whose solution is given by
\begin{gather} \label{n4}
g (u_{0,0},u_{0,1} )=\theta (\xi )+C\log (u_{0,0} )+D,\qquad \xi=u_{0,0}\left(u_{0,1}+\frac{1}{\epsilon}\right),
\end{gather}
where $C$ and $D$ are arbitrary constants and $\theta\not=0$ is an arbitrary function of its arguments.
Introducing (\ref{n4}) in (\ref{3.6aa}), (\ref{3.6aaab}) we get $b=\frac{c}{\alpha}$. The f\/inal determining equation (\ref{3.4aa}) implies $C=D=0$, $a=\alpha$ and $\theta=k u_{0,0} (1 + \epsilon u_{0,1})$. So, in conclusion (\ref{n1}) is linearizable by the transformation
\[ 
g (u_{0,0},u_{0,1} )= u_{0,0} (1 + \epsilon u_{0,1})
\]
into the linear equation
\[ 
\tilde u_{0,0} + \alpha \tilde u_{1,0} + b \tilde u_{0,1} + \alpha b \tilde u_{1,1}=0.
\]

\subsection[Four $QRT$-type linearizable equations]{Four $\boldsymbol{QRT}$-type linearizable equations}

As a f\/inal example we consider, as suggested by one of the referees,  the four $QRT$-type lineari\-zable nonlinear partial dif\/ference equations
recently presented in~\cite{sh}
\begin{gather*}
u_{1,1} = \frac{u_{1,0}+u_{0,1}-(1-u_{1,0}u_{0,1})u_{0,0}}{1-u_{1,0}u_{0,1}+(u_{1,0}+u_{0,1})u_{0,0}},\\ \nonumber
u_{1,1} = \frac{u_{1,0}-u_{0,1}+(1+u_{1,0}u_{0,1})u_{0,0}}{1+u_{1,0}u_{0,1}-(u_{1,0}-u_{0,1})u_{0,0}},\\ \nonumber
u_{1,1} = \frac{f_{2}(n,m)+f_{1}(n,m)u_{0,0}}{f_{1}(n,m)-f_{2}(n,m)u_{0,0}},\\ \nonumber
\nonumber f_{1}(n,m) \doteq (1+u_{1,0}u_{0,1})\big(u_{1,0}^2u_{0,1}^2-3u_{1,0}^2-3u_{0,1}^2+8u_{1,0}u_{0,1}+1\big),\\
\nonumber f_{2}(n,m) \doteq (u_{0,1}-u_{1,0})\big(3u_{1,0}^2u_{0,1}^2-u_{1,0}^2-u_{0,1}^2+8u_{1,0}u_{0,1}+3\big),\\ \nonumber
u_{1,1} = \frac{2u_{1,0}-u_{0,1}+u_{0,1}u_{1,0}^2-\big(1+2u_{1,0}u_{0,1}-u_{1,0}^2\big)u_{0,0}}{1+2u_{1,0}u_{0,1}-u_{1,0}^2
+\big(2u_{1,0}-u_{0,1}+u_{0,1}u_{1,0}^2\big)u_{0,0}}.
\end{gather*}
All these equations are linearizable through the transformation $v_{0,0}=\arctan (u_{0,0} )$ and give   the following linear equations
\begin{gather} \nonumber
v_{0,0}-v_{1,0}-v_{0,1}+v_{1,1}=p\pi,\\
v_{0,0}+v_{1,0}-v_{0,1}-v_{1,1}=p\pi,\label{Venus}\\ \nonumber
v_{0,0}-3v_{1,0}+3v_{0,1}-v_{1,1}=p\pi,\\ \nonumber
v_{0,0}-2v_{1,0}+v_{0,1}+v_{1,1}=p\pi,
\end{gather}
where $p\in\mathcal Z$. The $p$-dependent right-hand side is a consequence of the multi-valuedness of the $\arctan$ function. We can choose $p=0$ if we choose the principle branch of the  $\arctan$ function, i.e.~$\arctan (0 )=0$.

Applying the formulas contained in  Section~\ref{section2} it is immediate to show that the necessary conditions (\ref{3.13b}), (\ref{3.13c}) of linearizability through a point transformation are identically satisf\/ied. The integration of equation~(\ref{3.11a}) gives
\[
f (u_{0,0} )=A [\arctan (u_{0,0} )+B ],
\]
where $A\not=0$ and $B$ are arbitrary constants. The dif\/ferential consequences of equation (\ref{3.4a}) (which in this case will  have a constant right-hand side not necessarily equal to zero) can be identically satisf\/ied modulo the $QRT$-type equations if and only if
\begin{alignat*}{4}
& a=-1,\qquad && b=-1,\qquad && c=1, &\\
& a=1,\qquad && b=-1,\qquad && c=-1,&\\
& a=-3,\qquad && b=3,\qquad && c=-1, &\\
 & a=-2,\qquad && b=1,\qquad && c=1. &
\end{alignat*}
With these values of the coef\/f\/icients ($a,  b,  c$),  (\ref{3.4a}) is identically satisf\/ied modulo the correspon\-ding equations. In the f\/irst three cases its right hand side is equal to~$p\pi$ for arbitrary $B$ while in the fourth case it is equal to~$B+p\pi$. As $B$ is arbitrary, for the sake of simplicity we will choose $B=0$ in all cases.

\section{Conclusions}\label{section6}

In this paper we have considered the possibility of linearizing nonlinear partial dif\/ference equations def\/ined on a quad-graph by the use of point one, two points  and a Hopf--Cole transformation. Imposing the existence of such transformations we obtained for the equations  some necessary linearizability conditions and some dif\/ferential equations for the transformations. We have applied our results to some nonlinear partial dif\/ference equations, some of which already known to be linearizable, to test our procedure. In the case of the nonlinear equation~(\ref{Orma}), obtained as a result of a classif\/ication of multilinear equations on the quad-graph by the multiple scales expansion up to f\/ifth order, we have been able to show that all linearizability conditions considered here imply the constraint on the coef\/f\/icients~(\ref{Amor}).

In the case of the Hietarinta equation we have been able to f\/ind out a new linearizing Hopf--Cole transformation depending just on one function, up to our knowledge unknown.

In the verif\/ication of the examples presented in \cite{sh}, we discovered an accidental misprint in~(\ref{Venus}), as in the original article the signs of $v_{1,0}$ and $v_{0,1}$ are inverted.

A few problems are still open.
We have just considered  here the case of one point, two points and two points Hopf--Cole transformations but it could be interesting to consider in the future more general cases, maybe combining it with results on the integrability of the nonlinear equations. Moreover one would like to include nonautonomous transformations and lattice dependent linear equations.

\vspace{-1mm}

\subsection*{Acknowledgments}

LD and SC have been partly supported by the Italian Ministry of Education and Research, PRIN
 ``Continuous and discrete nonlinear integrable evolutions: from water
waves to symplectic maps'' from 2010.

\vspace{-3mm}

\pdfbookmark[1]{References}{ref}
\LastPageEnding


\begin{thebibliography}{99}

\footnotesize\itemsep=-1.0pt

\bibitem{abel}
Abel N.H.,
Methode generale pour trouver des fonctions d'une seule quantite variable lorsqu'une
propriete de ces fonctions est exprimee par une equation entre deux variables,
{\it Mag. Naturvidenskab.} {\bf 1} (1823), 1--10, reproduced in  Ouvres Completes, Vol.~I, Christiania, 1881, 1--10.\\
 Aczel  J.,
 Lectures on functional equations and their applications,
{\it Mathematics in Science and Engineering}, Vol.~19,
 Academic Press, New York~-- London, 1966.


\bibitem{cole}
Cole J.D.,
On a quasi-linear parabolic equation occurring in aerodynamics,
{\it Quart. Appl. Math.} {\bf 9} (1951), 225--236.


\bibitem{hls}
Hernandez Heredero R., Levi  D.,  Scimiterna C.,
A discrete linearizability test based on multiscale analysis,
\href{http://dx.doi.org/10.1088/1751-8113/43/50/502002}{{\it  J.~Phys.~A: Math. Theor.}}  {\bf 43} (2010), 502002, 14~pages,
\href{http://arxiv.org/abs/1011.0141}{arXiv:1011.0141}.


\bibitem{h04}
Hietarinta J.,
A new two-dimensional lattice model that is `consistent around a~cube',
\href{http://dx.doi.org/10.1088/0305-4470/37/6/L01}{{\it J.~Phys.~A: Math. Gen.}}  {\bf 37} (2004), L67--L73,
\href{http://arxiv.org/abs/nlin.SI/0311034}{nlin.SI/0311034}.

\bibitem{hopf}
Hopf E.,
The partial dif\/ferential equation $u_t + uu_x = u_{xx}$,
\href{http://dx.doi.org/10.1002/cpa.3160030302}{{\it  Comm. Pure Appl. Math.}}  {\bf  3} (1950), 201--230.


\bibitem{blr}
Levi D., Ragnisco  O., Bruschi M.,
Continuous and discrete matrix Burgers hierarchies,
\href{http://dx.doi.org/10.1007/BF02721683}{{\it Nuovo Cimento~B}} {\bf 74} (1983), 33--51.

\bibitem{ly09}
Levi  D., Yamilov R.I.,
The generalized symmetry method for discrete equations,
\href{http://dx.doi.org/10.1088/1751-8113/42/45/454012}{{\it J.~Phys.~A: Math. Theor.}}  {\bf 42} (2009), 454012, 18~pages,
\href{http://arxiv.org/abs/0902.4421}{arXiv:0902.4421}.

\bibitem{ly11}
Levi  D., Yamilov R.I.,
Generalized symmetry integrability test for discrete equations on the square lattice,
\href{http://dx.doi.org/10.1088/1751-8113/44/14/145207}{{\it J.~Phys.~A: Math. Theor.}} {\bf 44} (2011), 145207, 22~pages,
\href{http://arxiv.org/abs/1011.0070}{arXiv:1011.0070}.


\bibitem{mwx11}
Mikhailov  A.V.,  Wang J.P., Xenitidis P.,
Recursion operators, conservation laws and integrability conditions for dif\/ference equations,
\href{http://dx.doi.org/10.1007/s11232-011-0033-y}{{\it Theoret. and Math. Phys.}} {\bf 167} (2011), 421--443,
\href{http://arxiv.org/abs/1004.5346}{arXiv:1004.5346}.

\bibitem{miura}
Miura R.M.,
Korteweg--de Vries equation and generalizations. I.~A~remarkable explicit nonlinear transformation,
\href{http://dx.doi.org/10.1063/1.1664700}{{\it  J.~Math. Phys.}} {\bf 9} (1968), 1202--1204.\\
 Miura R.M., Gardner C.S., Kruskal M.D.,
 Korteweg--de Vries equation and generalizations. II.~Existence of conservation laws and constants of motion,
\href{http://dx.doi.org/10.1063/1.1664701}{{\it J.~Math. Phys.}} {\bf 9} (1968), 1204--1209.


\bibitem{jrg}
Ramani A., Joshi N.,  Grammaticos B.,  Tamizhmani T.,
Deconstructing an integrable lattice equation,
\href{http://dx.doi.org/10.1088/0305-4470/39/8/L01}{{\it J.~Phys.~A: Math. Gen.}} {\bf 39} (2006), L145--L149.

\bibitem{s11}
Startsev S.Ya.,
On non-point invertible transformations of dif\/ference and dif\/ferential-dif\/ference equations,
\href{http://dx.doi.org/10.3842/SIGMA.2010.092}{{\it SIGMA}} {\bf 6} (2010), 092, 14 pages,
\href{http://arxiv.org/abs/1010.0361}{arXiv:1010.0361}.



\bibitem{sh}
Uma Maheswari  C., Sahadevan R.,
On the conservation laws for nonlinear partial dif\/ference equations,
\href{http://dx.doi.org/10.1088/1751-8113/44/27/275203}{{\it  J.~Phys.~A: Math. Theor.}}  {\bf 44} (2011), 275203, 16~pages.

\end{thebibliography}
\end{document}